\documentclass{aa}
\usepackage[varg]{txfonts}

\usepackage{txfonts}

\usepackage[version=3]{mhchem}

\usepackage{natbib}
\usepackage[utf8]{inputenc} 
\usepackage[T1]{fontenc}    

\usepackage{graphicx}
\usepackage{afterpage}

\usepackage{amsmath}
\usepackage{diagbox}
\usepackage{multirow}

\usepackage{booktabs}
\usepackage{xcolor}

\begin{document}
\newcommand\nratio{{$^{14}$N/$^{15}$N}}
\newcommand{\rnow}{\ensuremath{{\cal R}_0}}
\newcommand{\rsun}{\ensuremath{{\cal R}_\odot}}
\newcommand{\rr}{\ensuremath{{\cal R}}}
\newcommand{\rrs}[1]{\ensuremath{{\cal R}_{\rm #1}}}
\newcommand{\joel}[1]{\textcolor{red}{[Joel notes: #1]}}

\title{Tracing Molecular Stratification within an Edge-on Protoplanetary Disk}  

\author{%
  D. Ru\'iz-Rodr\'iguez\inst{1,2} \and J. Kastner\inst{2} \and P. Hily-Blant\inst{3} \and T. Forveille\inst{4} } \institute{%
  National Radio Astronomy Observatory, 520 Edgemont Rd., Charlottesville, VA 22903, USA   \email{druiz@nrao.edu} \and Chester F. Carlson Center for Imaging Science, School of Physics \&
  Astronomy, and Laboratory for Multiwavelength Astrophysics,
  Rochester Institute of Technology, 54 Lomb Memorial Drive, Rochester
  NY 14623 USA \and Institut Universitaire de France \and Universit\'e
  Grenoble Alpes, CNRS, IPAG, F-38000 Grenoble, France}

\date{}

\abstract{High-resolution observations of edge-on proto-planetary disks in emission from molecular species sampling different critical densities and formation pathways offer the opportunity to trace the vertical chemical and physical structures of protoplanetary disks. Among the problems that can be addressed is the origin and significance of the bright CN emission that is a ubiquitous feature of disks. Based on analysis of sub-arcsecond resolution Atacama Large Millimeter Array (ALMA) archival data for the edge-on Flying Saucer disk (2MASS J16281370-2431391), we establish the vertical and radial differentiation of the disk CN emitting regions with respect to those of $^{12}$CO and CS, and we model the disk physical conditions from which the CN emission arises. We demonstrate that the disk $^{12}$CO (2-1), CN (2-1), and CS J=5-4 emitting regions decrease in scale height above the midplane, such that $^{12}$CO, CN, and CS trace layers of increasing density and decreasing temperature. We find that at radii $>$ 100 au from the central star, CN emission arises predominantly from intermediate layers, while in the inner region of the disk, CN appears to arise from layers closer to the midplane. We investigate disk physical conditions within the CN emitting regions, as well as the ranges of CN excitation temperature and column density, via RADEX non-LTE modeling of the three brightest CN hyperfine lines.  Near the disk midplane, where we derive densities $\rm n_{H_{2}}$ $\sim$10$^{7}$ cm$^{-3}$ at relatively low $\rm T_{kin}$ ($\sim$ 12 K), we find that CN is thermalized, while sub-thermal, non-LTE conditions appear to obtain for CN emission from higher (intermediate) disk layers. We consider whether and how the particular spatial location and excitation conditions of CN emission from the Flying Saucer can be related to CN production that is governed, radially and vertically, by the degree of irradiation of the flared disk by X-rays and UV photons from the central star.}


%



\maketitle

\section{Introduction}

Detailed chemical studies of dusty circumstellar (protoplanetary) disks orbiting young stars represent necessary steps toward understanding how exoplanets and exocomets obtain their chemical inventories. The unprecedented combination of sensitivity and spatial and spectral resolution afforded by the Atacama Large Millimeter Array (ALMA) is now enabling such investigations \citep[e.g.,][]{Qi2013, Oberg2015, Guzman2017, HilyBlant2017,Bergner2018,Kastner2018,Semenov2018, Qi2019,  Oberg2019}.

Among the problems that can be addressed via spatially resolved ALMA studies of protoplanetary disk chemistry is the origin and significance of the bright CN emission that is a ubiquitous feature of disks \citep[e.g.,][]{Kastner1997,Kastner2008,Kastner2014,Thi2004,Sacco2014,Guilloteau2014}. Different astrochemical models of protoplanetary disk CN production, most of which attempt to account for the influence of stellar high-energy irradiation, have made a variety of predictions as to CN abundances and locations \citep[e.g.][]{Lepp1996, Stauber2005, Chapillon2012, Cleeves2013, Cazzoletti2018}. In particular, most models predict that CN is mostly produced within disk surface layers via UV irradiation \citep[either through photodissociation of HCN or via vibrational excitation of H$_2$;][]{Cazzoletti2018,Visser2018}. However, the low CN kinetic temperatures inferred from observations \citep[$T \le 25$ K; e.g.][]{Kastner2014, Teague2016, HilyBlant2017} suggest that disk CN production may be driven by deeper-penetrating X-ray radiation from the central T Tauri stars \citep{Kastner2014}.

Confirming or ruling out whether CN can be formed deep within protoplanetary disks would have several interesting implications. Foremost, it would contribute to establishing whether CN is a photodissociation product of HCN, or is formed via other channels, such as reactions of N with C$_2$H \citep{Cazzoletti2018}. This, in turn, would help to constrain the physical and chemical processes that took place during the formation of the various isotopic reservoirs of nitrogen present in the Solar System \citep{Guzman2017,HilyBlant2017, HilyBlant2019}.

Despite the status of CN as a signature chemical tracer within disks --- and the foregoing open questions concerning its production and its utility as a tracer of UV irradiation and nitrogen isotopic ratios --- the only ALMA CN images of disks published thus far are those for the nearly pole-on TW Hya and a handful of moderately inclined disks \citep{Teague2016, HilyBlant2017,vanTerwisga2018, Teague2020}. The ring-like CN emission morphologies revealed by these images motivated recent model calculations by \citet{Cazzoletti2018} and, indeed, the observed morphologies are well reproduced by these simulations. However, the degeneracy inherent to low-inclination viewing geometries has left constraints on the scale height of CN emission entirely in the realm of the aforementioned temperature-based inferences. 

To more directly investigate the vertical distribution of CN molecules within protoplanetary disks, we have exploited archival ALMA data obtained for the nearly edge-on Flying Saucer disk (2MASS J16281370$-$2431391). The CN data analyzed here were obtained as part of the comprehensive study of the vertical and radial physical (density and temperature) structure of that disk carried out by \citet{Guilloteau2016}, and subsequently by \citet{Dutrey2017}.

The Flying Saucer system consists of a 0.57 M$_{\odot}$ young stellar object surrounded by a nearly edge-on disk ($i \approx 87^\circ$) of mass 3.5 $\times$ 10$^{-5}$ M$_{\odot}$. Its distance is estimated at $\sim$ 120 pc \citep{Loinard2008}. As no parallax is available for the Flying Saucer system in Gaia DR2 \citep{Gaia2018}, we adopt this distance for all calculations in this paper. The aforementioned ALMA observations revealed the $^{12}$CO J = 2-1 emission lying well above and below the disk plane, arising from upper and lower layers, extending to disk radii of at least 330 au. By contrast, the CS J = 5-4 molecular emission is confined towards the midplane, and extends to $\sim$ 300 au in radius \citep{Dutrey2017}. Contrary to the radial extension of the CS and $^{12}$CO (hereafter CO) emission, the dust continuum is limited to a more compact region with a radius of 190 au and, similarly to the CS emission, is confined towards the midplane, where the dust temperature is remarkably low \citep[$T_{Dust} \sim$ 5-8 K at 100 au;][]{Guilloteau2016}. Modeling of continuum and near-IR emission suggests that the disk is in an advanced state of dust settling, with scale heights of 22.5 $\pm$ 1.5 au for small grains ($\sim$ 1 $\mu$m) and 12.7 $\pm$ 0.3 au for large grains ($\sim$ 1 mm) at 100 au \citep{Grosso2003}.

Indeed, CO, CS, and 242 GHz continuum emission reveal a vertical stratification of the dust and gas in the system \citep{Dutrey2017}. These stratified disk density and temperature tracers hence potentially afford a direct means to constrain CN production pathways under different thermal and density conditions. To exploit this potential, we have analyzed the CN emission line data for the Flying Saucer disk that were obtained during the same ALMA observations that yielded the maps of CO, CS, and continuum \citep{Dutrey2017}. Sections \ref{Sec:Obs} and \ref{Sec:Results} present details of the observations, and the CS, CN, and CO molecular emission line maps resulting from our reprocessing of the archival ALMA data. In Sec. \ref{Sec:Results}, we present an analysis of the disk vertical intensity distributions and CN emission line profiles, together with the derivation of the disk physical parameters. Section \ref{Sec:Discussion} presents a discussion of the disk structure, CN abundances, and potential triggers of CN production in the disk at different scale heights and radii.

\begin{table*}
\centering
\small
\caption{Detected CN (N = 2-1) HF transitions.}
  \begin{tabular}{cccccc}
  \toprule
    
 \textbf{HF Component}$^{a}$  &  \textbf{Transition}  &\textbf{ $\nu$} & A$_{ul}$& g$_{u}$ & Relative Intensities$^{b}$\\
   
 &   &[GHz] & & &  \\
    
     \cline{1-6}
A&    J = 5/2 $\rightarrow$ 3/2, F = 5/2 $\rightarrow$ 3/2 &226.8741908  & 9.62$\times$10$^{-5} $ & 6 & 0.1805  \\ 
B &     J = 5/2 $\rightarrow$ 3/2, F = 7/2 $\rightarrow$ 5/2 & 226.8747813  & 1.14$\times$10$^{-4} $   & 8 & 0.2860\\ 
 C&     J = 5/2 $\rightarrow$ 3/2, F = 3/2 $\rightarrow$ 1/2 &226.8758960  & 8.59$\times$10$^{-5} $  & 4 & 0.1074\\ 
 D&    J = 5/2 $\rightarrow$ 3/2, F = 3/2 $\rightarrow$ 3/2 &226.8874202  & 2.73$\times$10$^{-5} $   & 4 &  0.0342 \\ 
E &    J = 5/2 $\rightarrow$ 3/2, F = 5/2 $\rightarrow$ 5/2 & 226.8921280  & 1.81$\times$10$^{-5} $   & 6 & 0.0340\\ 
    \hline
  \end{tabular}
  \label{Table:Transitions}
  
\begin{flushleft}
$^{a}$ CN (2-1) has 19 HF components. We list here the 5 brightest (hence well-detected) HF components within the observed bandpass, referred to throughout this paper as A-E. Molecular data from \citet{Skatrud1983}. \\
$^{b}$ Theoretical relative intensities under LTE conditions of the five brightest HF transitions of CN within the bandwidth, adapted from Table 2 of \citet{Punzi2015}.\\
\end{flushleft}  
 \end{table*}

\section{Observations}
\label{Sec:Obs}

\begin{figure*}
\centering
\includegraphics[width=0.49\textwidth]{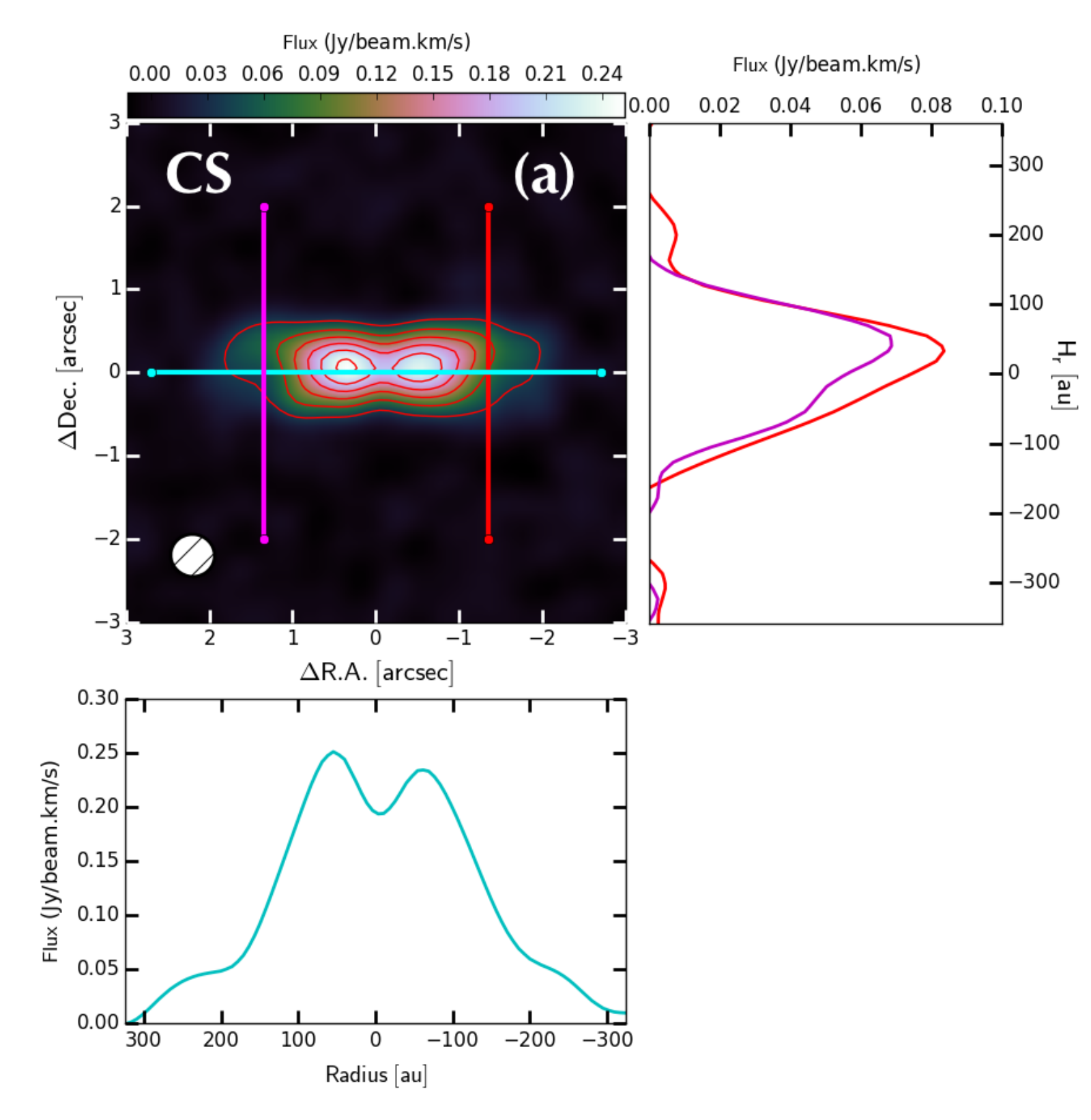}
\includegraphics[width=0.49\textwidth]{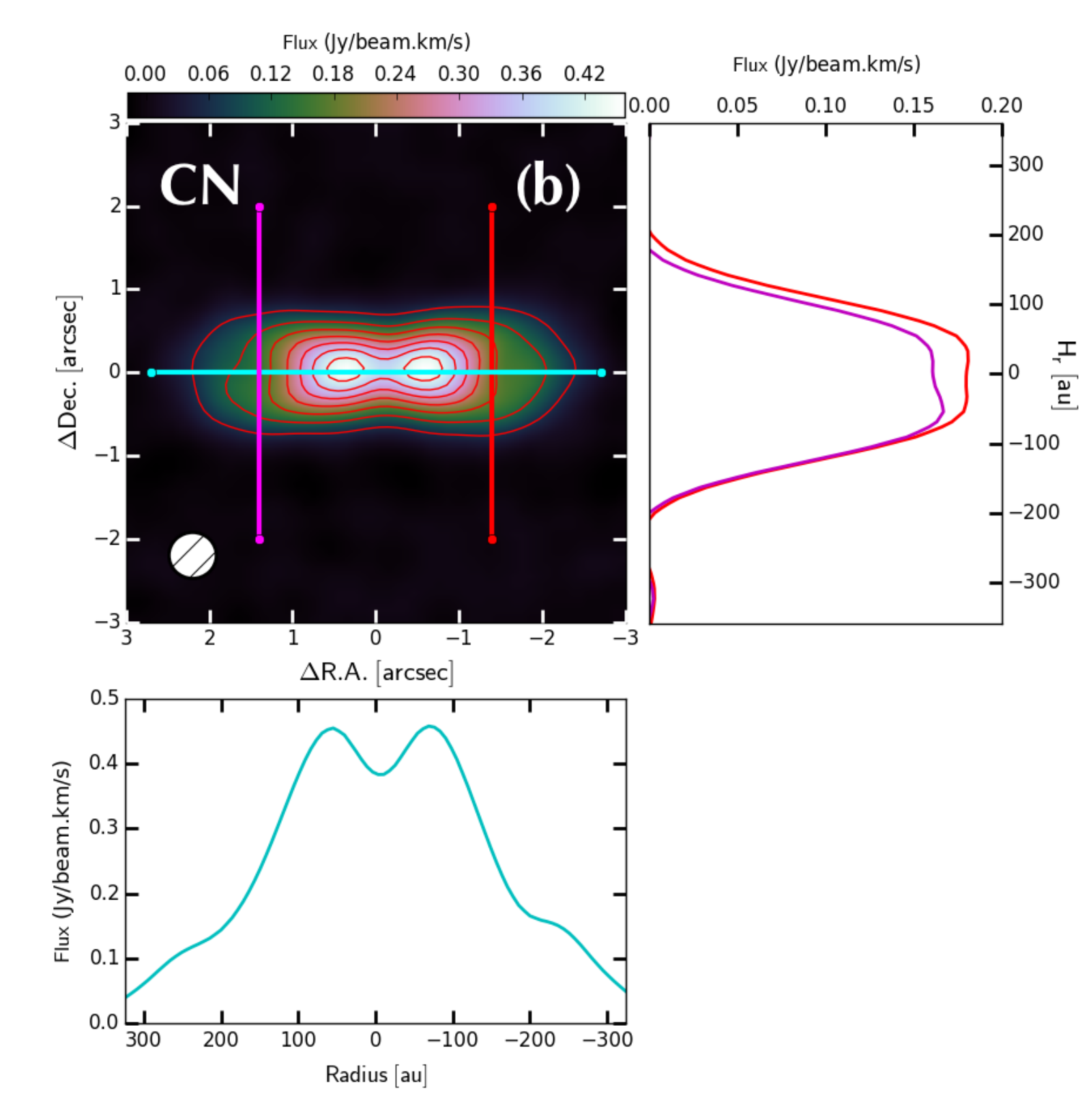}
\includegraphics[width=0.49\textwidth]{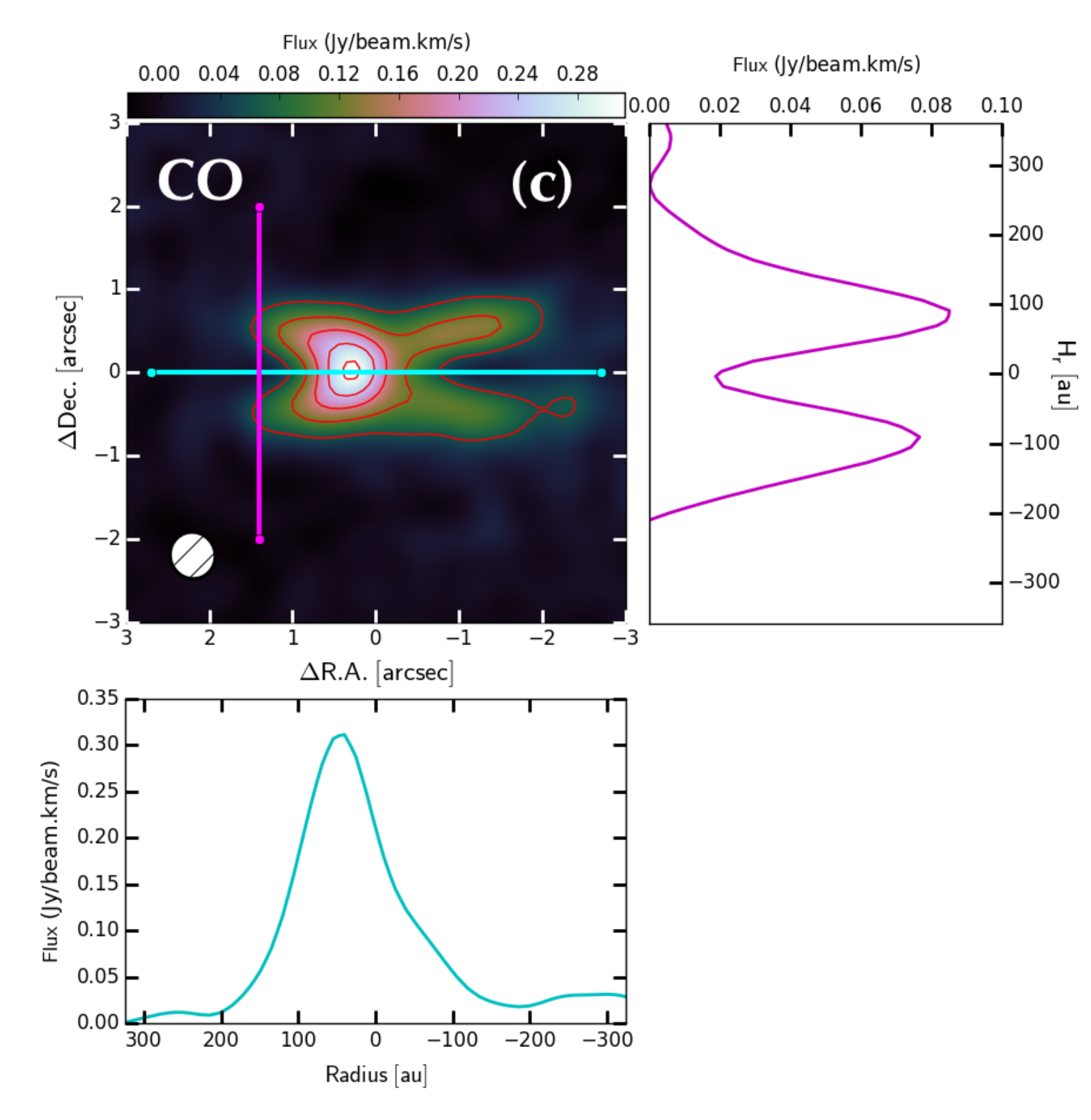}
\includegraphics[width=0.49\textwidth]{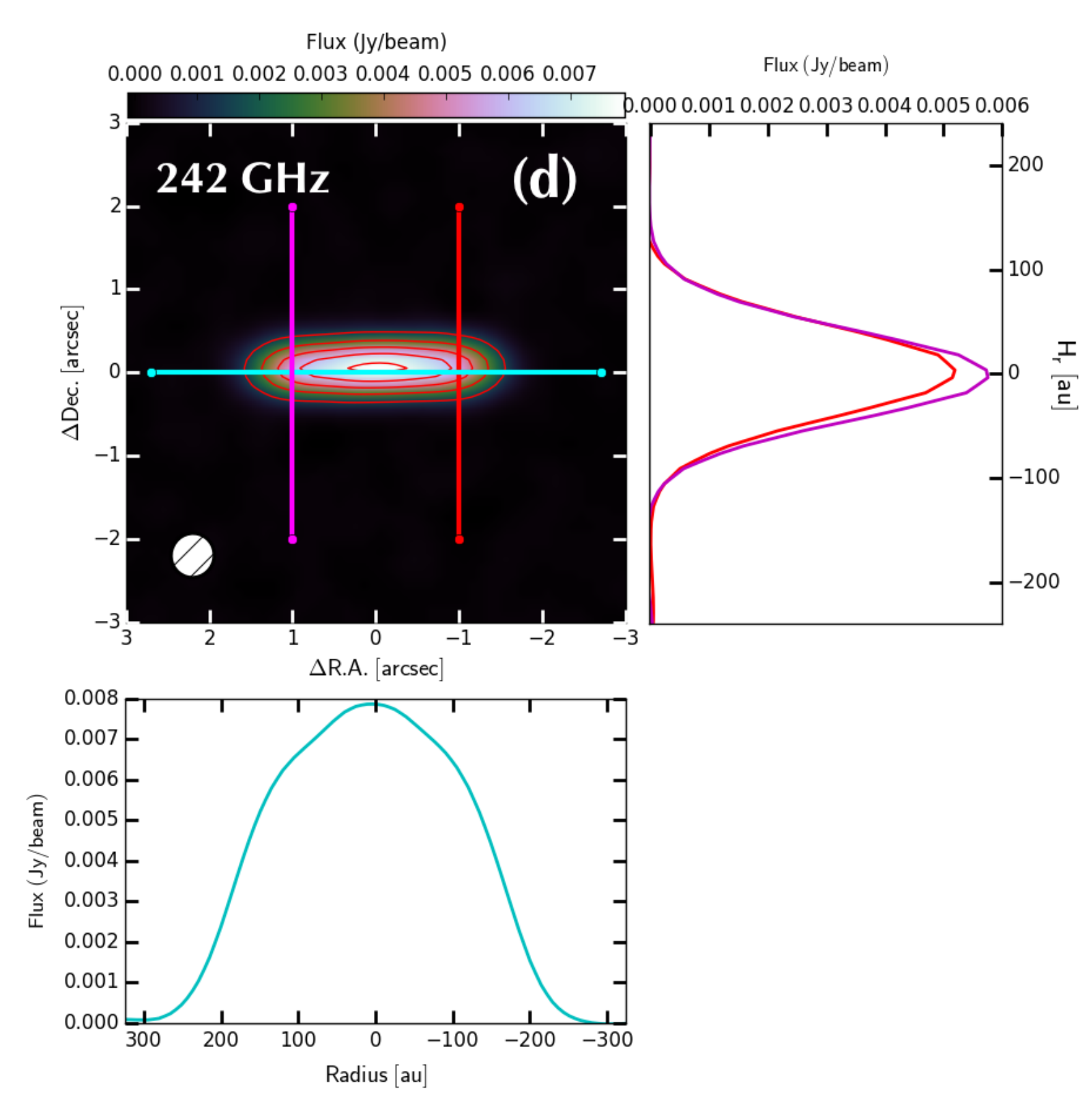}
\caption{(a) Integrated emission map of CS from -0.9 to 8.6 km s$^{-1}$ with contours showing 10,  20,  30,  40,  50, and 60 $\sigma$ emission ($\sigma$ = 55 mJy beam$^{-1}$ km s$^{-1}$); the 0.53$^{''}$$\times$ 0.51$^{''}$ at P.A. -57.8$^{\rm o}$ beam is indicated in the bottom left corner. (b) Integrated emission map of CN from 0.8 to 10.4 km s$^{-1}$ with contours showing 10,  20,  30,  40,  50,  60,  70, and 80 $\sigma$ emission $\sigma$ = 60 mJy beam$^{-1}$ km s$^{-1}$, and a 0.57$^{''}$$\times$ 0.55$^{''}$ at P.A. -55.0$^{\rm o}$ beam. (c) Integrated emission map of CO from -0.9 to 8.6 km s$^{-1}$ with contours showing 10,  20,  30, and 40 $\sigma$ emission ($\sigma$ = 90 mJy beam$^{-1}$ km s$^{-1}$), and a 0.56$^{''}$$\times$ 0.54$^{''}$ at P.A. -63.4$^{\rm o}$ beam. (d) ALMA 1.236 mm dust continuum image (natural weighting) with a 0.64$^{''}$$\times$ 0.60$^{''}$ at P.A. -56$^{\rm o}$ beam. At the bottom of each panel are shown the radial intensity profiles of the integrated CS, CN and CO emissions and 242 GHz continuum. The radial profiles are extracted along a cut through the disk midplane (P.A.= 3$^{\rm o}$) and displayed by cyan lines in each image. Similarly, at the right of each panel are shown the representative vertical intensity profiles of the integrated CS, CN and CO emissions. The vertical profiles are extracted along a cut at a radius of 170 au for CS, CN, and CO, and at 120 au for the 242 GHz continuum. See Section \ref{Sec:Scale}.
\label{Fig:cuts}}
\end{figure*}

The CN N=2-1, CS J = 5-4 and CO J = 2-1 observations of the Flying Saucer presented in this work were obtained on May 17 and 18, 2015 as part of the Cycle 2 program 2013.1.00387.S (PI: Guilloteau). Using 37 twelve meter antennas, the range of unprojected baseline length is from 15 m to 560 m, and the total observation time is $\sim$ 55 minutes. As previously noted, a detailed description and analysis of the continuum, CO, and CS data was presented in \citet{Guilloteau2016} and \citet{Dutrey2017}.

Here, we reprocessed the ALMA data. To that end, the ALMA visibility data were edited, calibrated and imaged using the pipeline version r39732 in CASA 4.2.2. To image the data, we used the TCLEAN algorithm. We performed self-calibration on the source continuum emission with three rounds of phase calibration. To achieve a good balance between sensitivity and angular resolution the Briggs weighting parameter $R$ was set to 0.5 for the continuum map and spectral line image cubes. For the continuum map this yields a synthesized beam of 0.53$^{''}$$\times$ 0.52$^{''}$ at P.A. -65.1$^{\rm o}$. The peak 242 GHz continuum flux is 8.5 mJy beam$^{-1}$ and the r.m.s. is 0.2 mJy beam$^{-1}$. Given the relatively weak continuum emission, continuum subtraction should not remove significant line flux \citep{Weaver2018}. For the lines of interest, we then subtracted the continuum emission by fitting a first-order polynomial to the continuum in the $uv$-plane, and produced spectral data cubes with 0.06 km s$^{-1}$ velocity resolution. The image cubes were constructed on a 256 $\times$ 256 pixel grid with 0.1$''$ pixel size. For CS, the resulting rms was $\sim$2 mJy beam$^{-1}$ ($\sim$ 0.10 K) with a synthesized beam of 0.53$^{''}$$\times$ 0.51$^{''}$ ($\sim$ 60 au) at P.A. -57.8$^{\rm o}$. For CO, the rms is $\sim$2.5 mJy beam$^{-1}$ ($\sim$ 0.14 K) with a synthesized beam of 0.56$^{''}$$\times$ 0.54$^{''}$ ($\sim$65 au) at P.A. -63.4$^{\rm o}$. For CN, the rms was $\sim$2 mJy beam$^{-1}$ ($\sim$ 0.11 K) with a synthesized beam of 0.57$^{''}$$\times$ 0.55$^{''}$ ($\sim$65 au) at P.A. -55.0$^{\rm o}$. In the latter case, the five brightest (of nineteen) CN (2-1) hyperfine (HF) components are detected within the frequency range covered by these observations, 226.87 - 226.89 GHz. These CN HF lines at 226.8741908, 226.8747813, 226.8758960, 226.8874202, 226.8921280 GHz are labeled as `A', `B', `C', `D' and `E' in Table \ref{Table:Transitions}, respectively.

We constructed moment 0 maps from each of the three spectral line data cubes (Sec.~\ref{Sec:Scale}), including a map merging CN components A--E. by clipping the signal at $<$1$\sigma_{rms}$. The total continuum flux is 37.4 $\pm$ 0.88 mJy at 242 GHz, which is in good agreement with 35 mJy with $\sim$7$\%$ calibration uncertainty obtained by \citet{Dutrey2017} from the same ALMA dataset. The integrated emission line intensities over the entire emitting region are 1.60 $\pm$ 0.05  Jy km s$^{-1}$, 3.90 $\pm$ 0.12  Jy km s$^{-1}$, and 1.80 $\pm$ 0.16 Jy km s$^{-1}$ for CS, CN, and CO, respectively. These flux uncertainties are estimated with the sum in quadrature of  $\lesssim$10$\%$ flux and rms (per the ALMA Cycle 2 Technical Handbook\footnote{https://almascience.eso.org/documents-and-tools/cycle-2/alma-technical-handbook/}). The CN line flux is consistent  with that previously obtained by \citet{Reboussin2015}, 3.4 $\pm$ 0.5 Jy km s$^{-1}$, with the IRAM  30 m telescope.

\section{Results}
\label{Sec:Results}

\subsection{Integrated Intensity Maps and Scale Heights}
\label{Sec:Scale}

\begin{table*}
\centering
\caption{Results of vertical profile analysis.}
  \begin{tabular}{ccccc}
  \toprule 
    {\textbf{Data}} &
    {\textbf{Radius}} 
    & \textbf{FWHM}$^{a}$ & \textbf{Beam} & \textbf{H$_{r}$}$^{c}$  \\
   
    &[au]  & [au] & [au] & [au] \\
     \cline{1-5}
     
       \textbf{242 GHz} & & 99.4 $\pm$ 0.3 & 63.8 & 32.0 $\pm$  1.2  \\     
      \textbf{CS} & &129.5 $\pm$ 0.9 &  63.3 &48.0 $\pm$ 0.7  \\      
   \textbf{CN}& 60 &   154.9 $\pm$ 0.7  & 68.2 & 59.1 $\pm$ 0.6 \\    
   \textbf{CO } & & 193.2 $\pm$ 1.7 & 67.5  & 76.9 $\pm$ 0.9 \\
    
    \hline
    
    \textbf{242 GHz} & & 99.7 $\pm$ 0.3 & 63.8 & 32.3 $\pm$  1.1  \\    
    \textbf{CS } &  &  143.5 $\pm$ 1.8 &  63.3 &54.7 $\pm$ 0.9  \\   
    \textbf{CN }& 100 &   169.1 $\pm$ 1.5 & 68.2 & 65.7 $\pm$ 0.8\\     
   \textbf{CO} & &  127.6 $\pm$ 3.0$^{b}$ & 67.5 &  107.2 $\pm$ 1.6  \\

    \hline
    
     \textbf{242 GHz} & & 103.4 $\pm$ 0.7 & 63.8 & 38.1 $\pm$ 1.2\\   
    \textbf{CS} &  &  159.0 $\pm$ 4.8 & 63.3 & 61.9 $\pm$ 2.3  \\ 
    \textbf{CN}& 180 &   205.7 $\pm$ 4.5 &  68.2 &82.4 $\pm$ 2.0  \\ 
   \textbf{CO } & &  123.6 $\pm$ 3.9$^{b}$ & 67.5 & 138.3$\pm$ 2.2   \\
    \hline
  \end{tabular}
  \label{Table:Profiles}
  
  \begin{flushleft}
  $^{a}$ Parameters obtained from Gaussian fit. For details, see Sec. \ref{Sec:Scale}.\\
  $^{b}$ Multi-gaussian fitting performed; see text.\\
  $^{c}$ Characteristic emission height obtained after deconvolving FWHM from beamsize.\\

\end{flushleft}  
\end{table*}

\begin{figure}
\centering
\includegraphics[width=0.48\textwidth]{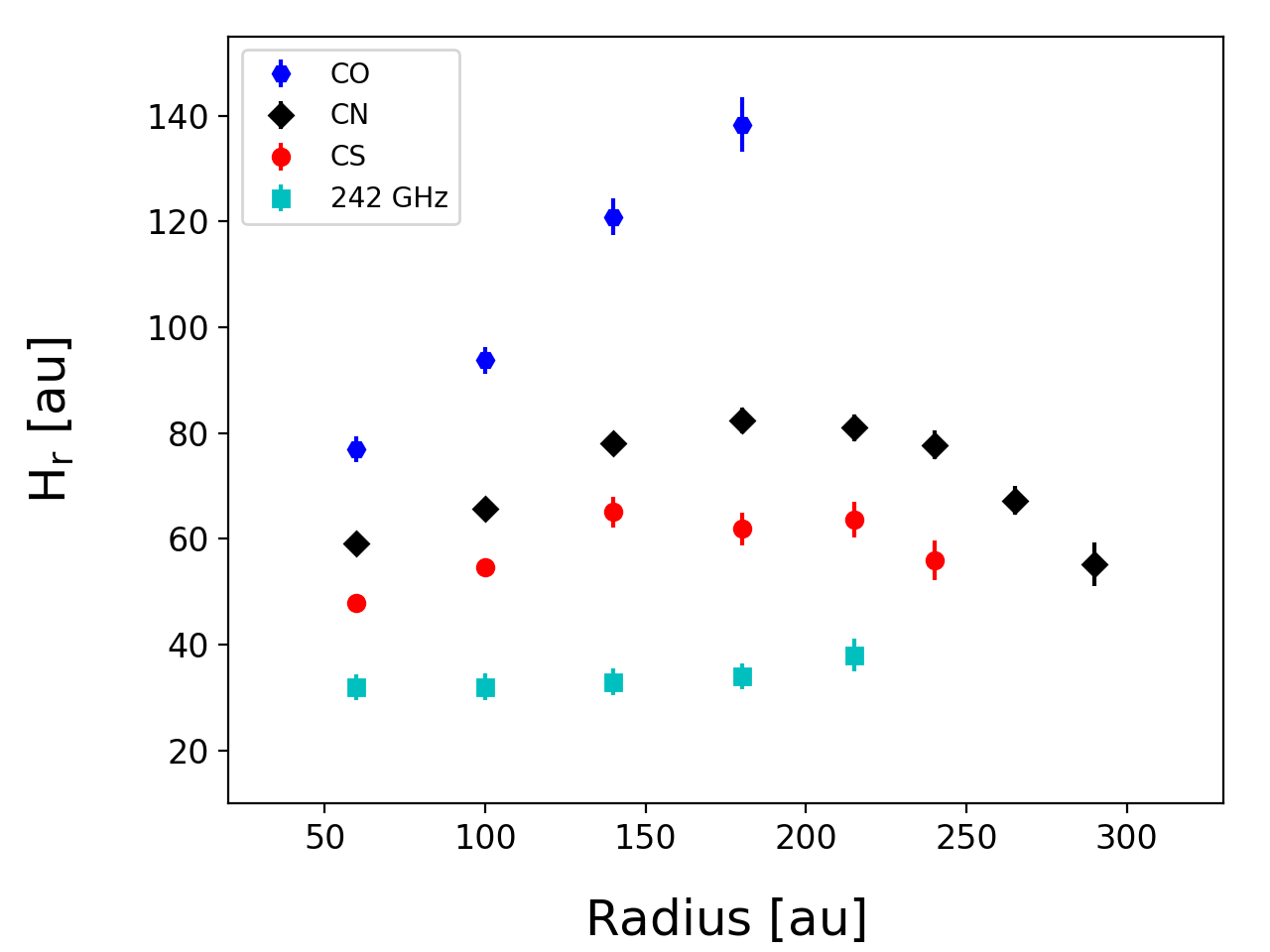}
\caption{Vertical scale heights estimated at various radii from the CO (blue hexagons), CN (black diamonds), CS (red dots), and 242 GHz continuum (cyan squares) integrated emission images. $H_{r}$ values are only measured over the area where the signal is $> 3\sigma$, while error bars display the standard error of the mean.
\label{Fig:layers_radii}}
\end{figure}

Figure \ref{Fig:cuts} shows the integrated intensity (moment 0) maps of the Flying Saucer disk in the CS J = 5-4, CN N=2-1, and CO J = 2-1 emission lines together with the 242 GHz continuum emission. Figure \ref{Fig:cuts} also displays CS, CN and CO radial intensity profiles along the midplane of the nearly edge-on disk in the images and representative vertical intensity profiles extracted along a cut at 170 au from the central source. In the  CN moment 0 image and radial intensity profile, the emission displays a deficit toward the central star, suggestive of an edge-on ring structure. The CS moment 0 map and its radial profile display a very similar, double-peaked morphology \citep[see also][]{Dutrey2017}. The western regions of the CO moment 0 image are adversely affected by contaminating background emission \citep[][]{Dutrey2017}, such that the radial profile of CO emission is difficult to interpret. Emission from both CN and CS peaks at radii of $\sim$70 au.

From the radial profiles, we find that the bulk of the CS line emission ($>$75$\%$) originates within approximately $\sim$200 au from the central source and has a total extension of $\sim$300 au, as also reported by \citet{Dutrey2017}. By contrast, the CN emitting region extends to at least 360 au, well beyond the maximum radius of CS emission. Moreover, at distances greater than 180 au, where the CO emission intensity in the mid-plane drops abruptly on the (uncontaminated) eastern side of the disk, the CN emission in the moment 0 map remains bright and confined closer to the mid-plane. 

To describe the molecular emission structure of the disk, we estimate the characteristic vertical disk height of CS, CN, and CO emission ($H_{r}$) at various radii from 60 au to $\sim$300 au. These vertical profiles were extracted along a position angle (PA) of $\sim$3.6$^\circ$ measured east from north \citep{Dutrey2017}. For the continuum, CS, and CN, which are marginally resolved we use the FWHMs of Gaussian fits to the observed vertical profiles to quantify $H_{r}$. 

Assuming that the characteristic emission height at a given radius $r$ is similar on the western and eastern sides of the disk,
the estimated FWHM values are averaged. Figure \ref{Fig:cuts}a and \ref{Fig:cuts}b illustrate representative profiles along vertical cuts at a radius of 170 au on each side, verifying their similarity. 
To obtain $H_{r}$, we then deconvolve the beam width ($\theta_{FWHM}$) from the resulting FWHM, i.e.,
\begin{equation}
\rm H_{r} = \frac{\left ( \sqrt{ \rm FWHM^{2} - \theta_{ \rm FWHM}^{2}} \right )}{2\sqrt{ \rm 2\: \textup{ln}2}}.
\label{Eq:Hr}
\end{equation}

Note that the resulting characteristic emission height $H_{r}$ is likely directly related to the vertical scale height of the molecular species in question. However, given the caveats, in particular, the emission from a given species only directly traces the disk density structure and abundance gradients if optically thin --- it is important not to equate the results for $H_{r}$ to disk scale height measurements.

In the specific case of CO emission at $r \geq$ 100, whose vertical extent is clearly larger than CS and CO, the shapes of the vertical profiles change considerably as a function of $r$, as a result of CO freeze-out processes and self-absorption by colder gas along the disk midplane \citep{Dutrey2017}. These effects are manifested in the form of a lack of CO emission from the midplane, generating double-peaked vertical intensity profiles (see Figure \ref{Fig:cuts}c). Hence, the disk scale height cannot be characterized by the above approximation (Eq. \ref{Eq:Hr}). Instead, the height of the disk surface can be estimated by accounting for the resulting double-peaked vertical profiles (see Figure \ref{Fig:cuts}c), whose peak positions ($\mu$) increase as disk radius increases in a flared disk. Thus, the upper and lower disk layers, also labeled as $H_{r}$, can be computed from
\begin{equation}
\rm H_{r} = \frac{\left ( \sqrt{ \rm FWHM^{2} - \theta_{FWHM}^{2}} \right )}{2\sqrt{2\: \textup{ln}2}}  +  \left |  \mu \right | ,
\label{Eq:Hr_co}
\end{equation}
with $ \rm FWHM$ defined as
\begin{equation}
\rm FWHM = 2\sqrt{ \rm 2\: \textup{ln}2}\sigma ,
\end{equation}

where $\mu$, and $\sigma$ are the peak position and the width of the Gaussian fitted, respectively. Here, $\mu$ accounts for an offset in the peak of the distribution. We then adopt the average of these (upper and lower) $H_{r}$ values. Bearing in mind that CO is contaminated by background emission from extended molecular clouds at velocities around $\sim$1.8 km s$^{-1}$ and in the range 6-7 km s$^{-1}$ \citep{Guilloteau2016}, we only analyze vertical profiles from the least contaminated region, i.e. the east side of the disk.  Overall, $H_{r}$ values are only measured over the area where the signal is $> 3\sigma$. Table \ref{Table:Profiles} displays the resulting $H_{r}$ values at 60, 100 and 200 au for the CS, CN and CO molecular lines. The tabulated uncertainties only reflect the formal errors on the Gaussian fits to the intensity profiles \footnote{We find very similar values of $H_{r}$ for CS and CN if we adopt the double-Gaussian fitting approach used for CO, i.e., if we use Equation \ref{Eq:Hr_co} rather than Equation \ref{Eq:Hr} to define $H_{r}$.}.

In Figure \ref{Fig:layers_radii}, we illustrate $H_{r}$ as a function of radius in the disk. It is immediately apparent that the peak emission from the three molecular species is stratified, with CO, CN, and CS lying increasingly close to the disk midplane (as traced by the dust continuum emission). We discuss these results in more detail in Sec \ref{Sec:Gradient}.

\subsection{CN Emission}
\label{Sec:CNemission}

\begin{figure*}
\centering
\includegraphics[width=0.9\textwidth]{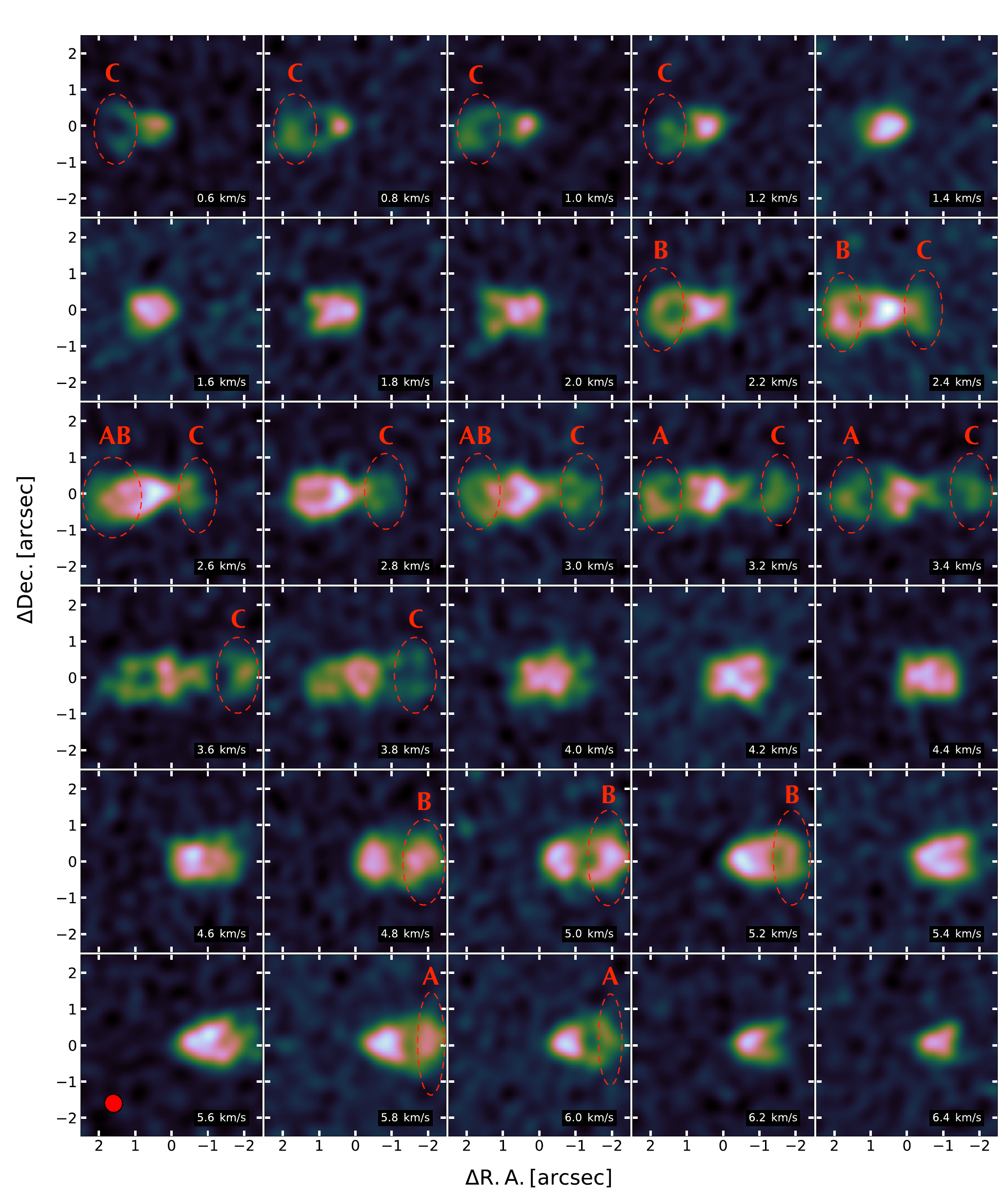}
\caption{Channel maps of the CN emission in the Flying Saucer from 0.60 to 6.40 km s$^{-1}$ with the velocity labels referring to the B component. The spectral resolution is 0.2 km s$^{-1}$ and the rms per channel is 14 mJy beam$^{-1}$.The synthesized beam size is represented in the lower left panel. The frequency
 range covers the `A', `B, and `C' HF components, which are J = 5/2 $\rightarrow$ 3/2, F = 5/2 $\rightarrow$ 3/2,  J = 5/2 $\rightarrow$ 3/2, F = 7/2 $\rightarrow$ 5/2, J = 5/2 $\rightarrow$ 3/2, F = 3/2 $\rightarrow$ 1/2, respectively, and listed in Table \ref{Table:Transitions}.
\label{Fig:Channel}}
\end{figure*}

Figure \ref{Fig:Channel} presents channel maps of CN emission, where the A, B, C HF emission components, indicated by the circles, are spatially resolved and appear according to their frequency-equivalent velocity. These CN channel maps were generated with central velocities from $+$0.8 to $+$6.6 km s$^{-1}$ with respect to the brightest HF component (labeled B in Table \ref{Table:Transitions}).

As is the case for the CO emission from the Flying Saucer \citep{Dutrey2017}, these CN channel maps reveal the vertical structure of a flared, gaseous disk, with blue-shifted material to the west (W) and redshifted material to the east (E). However, in these CN channel maps the HF splitting of the CN molecule visually distorts the typical emission expected from Keplerian rotation seen at high inclinations.  This visual distortion is caused by the overlapping of the HF components in velocity space at different locations in the disk, as is apparent from the PV diagram presented in Fig. \ref{Fig:PV}. Furthermore, the disk shows nearly symmetric CN emission, with an emission deficit in the midplane at radii between $\sim$ 80 and 200 au. The CN emission from this region seems to trace upper and lower surfaces, which, in some iso-velocity curves, reconnect at the outer region of the disk, forming an outer wall. This emission feature --highlighted by the B component-- is most evident in channels at projected velocities of $\approx$ 2.4 and  $\pm$ 0.2 km s$^{-1}$ and 5.0 and  $\pm$ 0.2 km s$^{-1}$ ($\approx$ 1.5 km s$^{-1}$ from the systemic velocity). A similar pattern is detected in the weaker components, D and E, which fall outside the range plotted in Fig.~\ref{Fig:Channel} (see Appendix \ref{App:A}), and in the channel maps of CS emission (Fig. \ref{Fig:CS_Channel}).

\subsubsection{Position-Velocity Diagram}
\label{Sec:PV}

\begin{figure}
\centering
\includegraphics[width=0.47\textwidth]{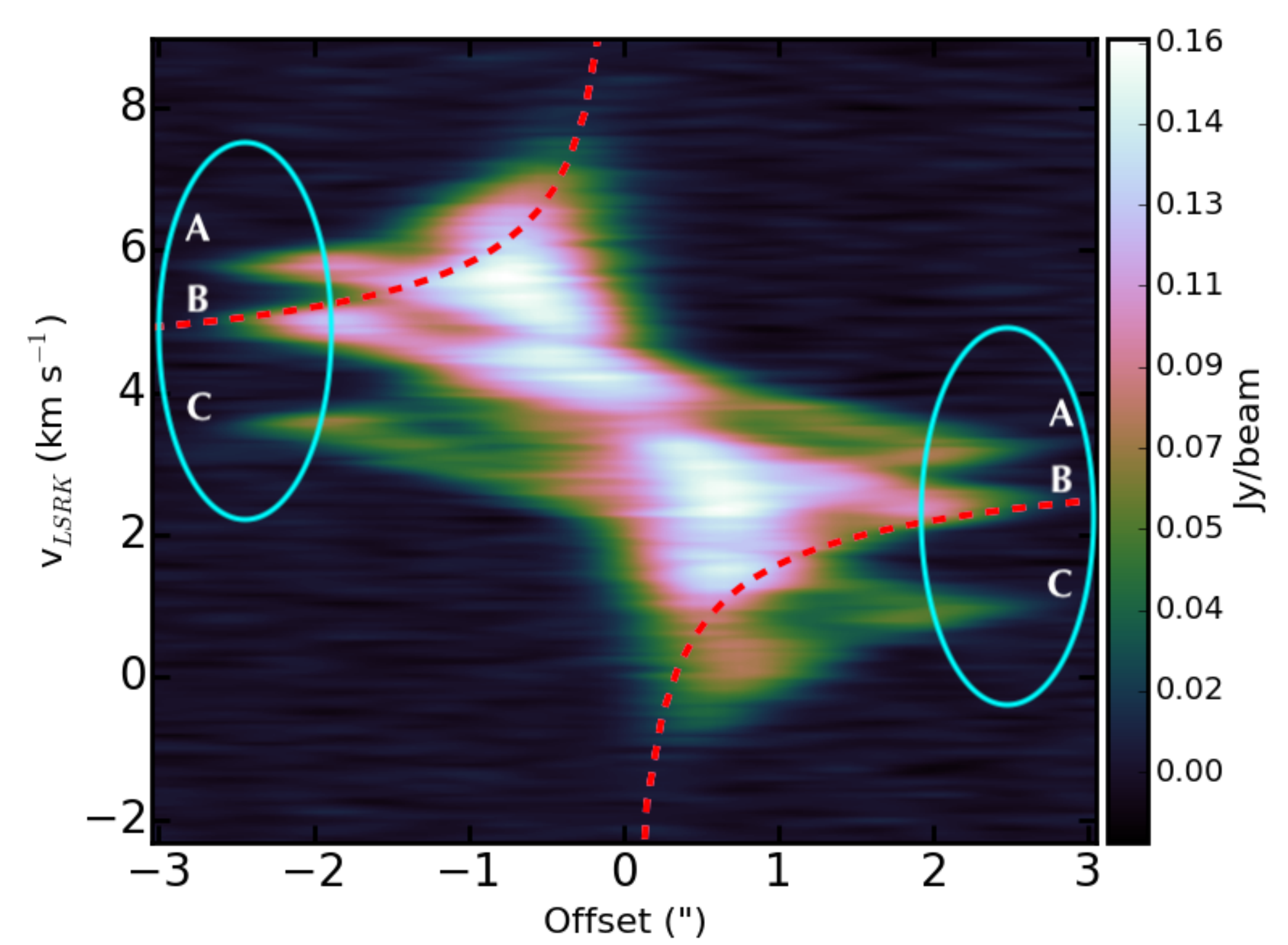}
\caption{Position-velocity diagram along the disk major axis of the CN emission. Dashed curve is a Keplerian  velocity  profile for a stellar mass of 0.61 M$_{\odot}$ at an inclination of 87$^{\rm o}$ and a systemic velocity of 3.70 km s$^{-1}$ LSRK. The V$_{LSRK}$ scale is with respect to the brightest HF component (B) detected at frequency 226.8747813 GHz. Labels A, B, and C indicate the brightest HF transitions, see Table \ref{Table:Transitions}. Cyan ellipses correspond to the regions from which we extracted the spectra analyzed in Sec. \ref{Sec:Profile}.
\label{Fig:PV}}
\end{figure}

Figure \ref{Fig:PV} displays a PV diagram centered on the three brightest CN (2-1) HF components (A, B, and C; Table \ref{Table:Transitions}), obtained from a cut along the major axis of the disk (i.e. P.A $\sim$ 93.5 deg.). In these as well as the fainter HF components (D and E; see Appendix \ref{App:A}), the PV diagram reveals a disk consistent with Keplerian rotation -- with redshifted emission to the east and blueshited to the west -- as is the case for the CO and CS lines \citep{Dutrey2017}. Taking the brightest HF line (B component) as the reference frame, the PV diagram is well reproduced by a Keplerian rotation profile in a geometrically thin disk with an inclination of 87 $\pm$ 1.0 deg., a central mass of 0.60 $\pm$ 0.02 M$_{\odot}$ and a Kinematic Local Standard of Rest velocity (v$_{LSRK}$) of 3.70 $\pm$ 0.5 km s$^{-1}$, assuming a distance of 120 pc. These values are in good agreement with the previous estimate of 0.58$\pm$ 0.01 M$_{\odot}$ obtained from a model of the disk CS emission by \citet{Dutrey2017}. 


\subsubsection{CN Line Profiles and Parameters}
\label{Sec:Profile}

\begin{figure}
\centering
\includegraphics[width=0.49\textwidth]{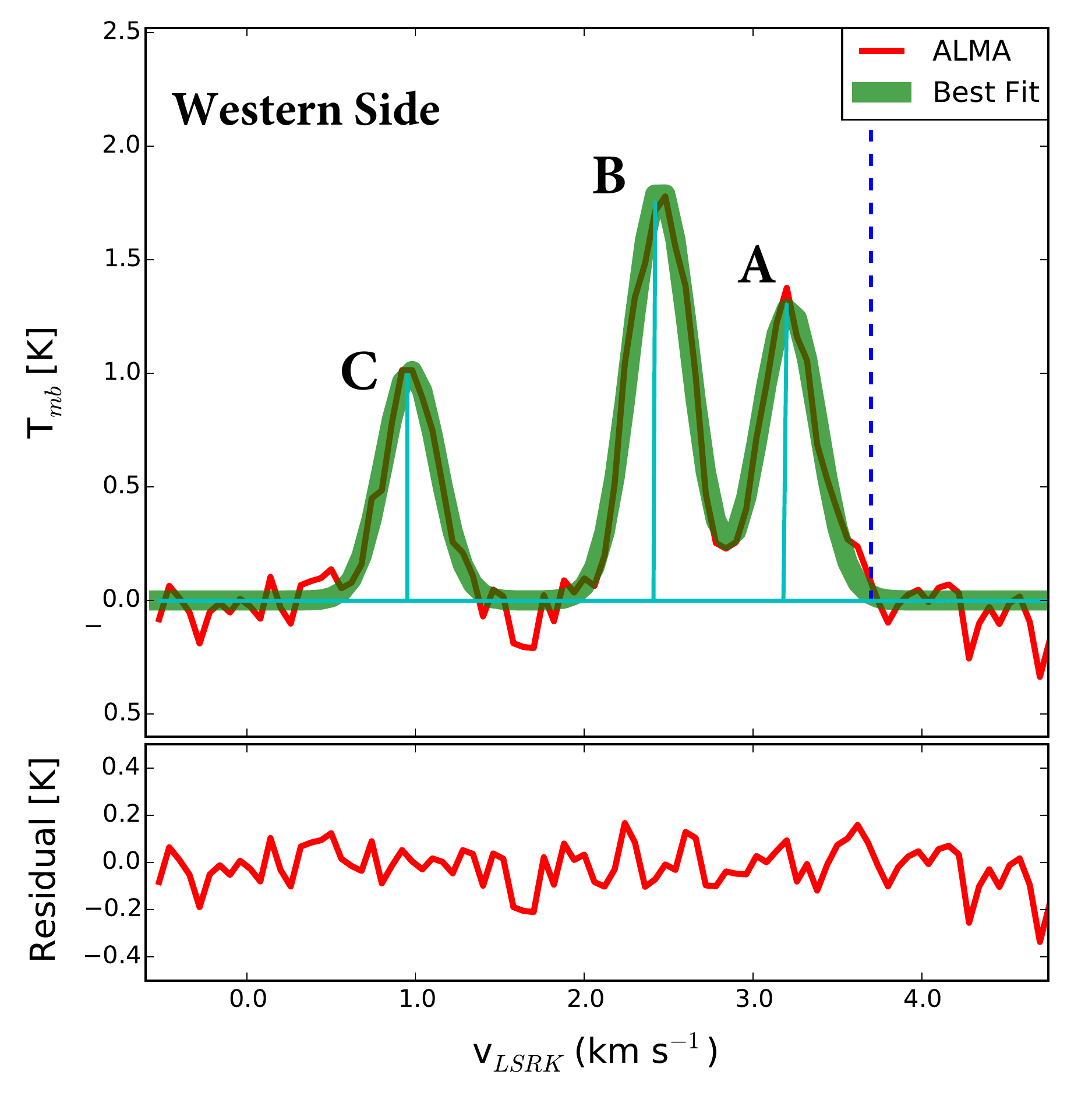}
\includegraphics[width=0.49\textwidth]{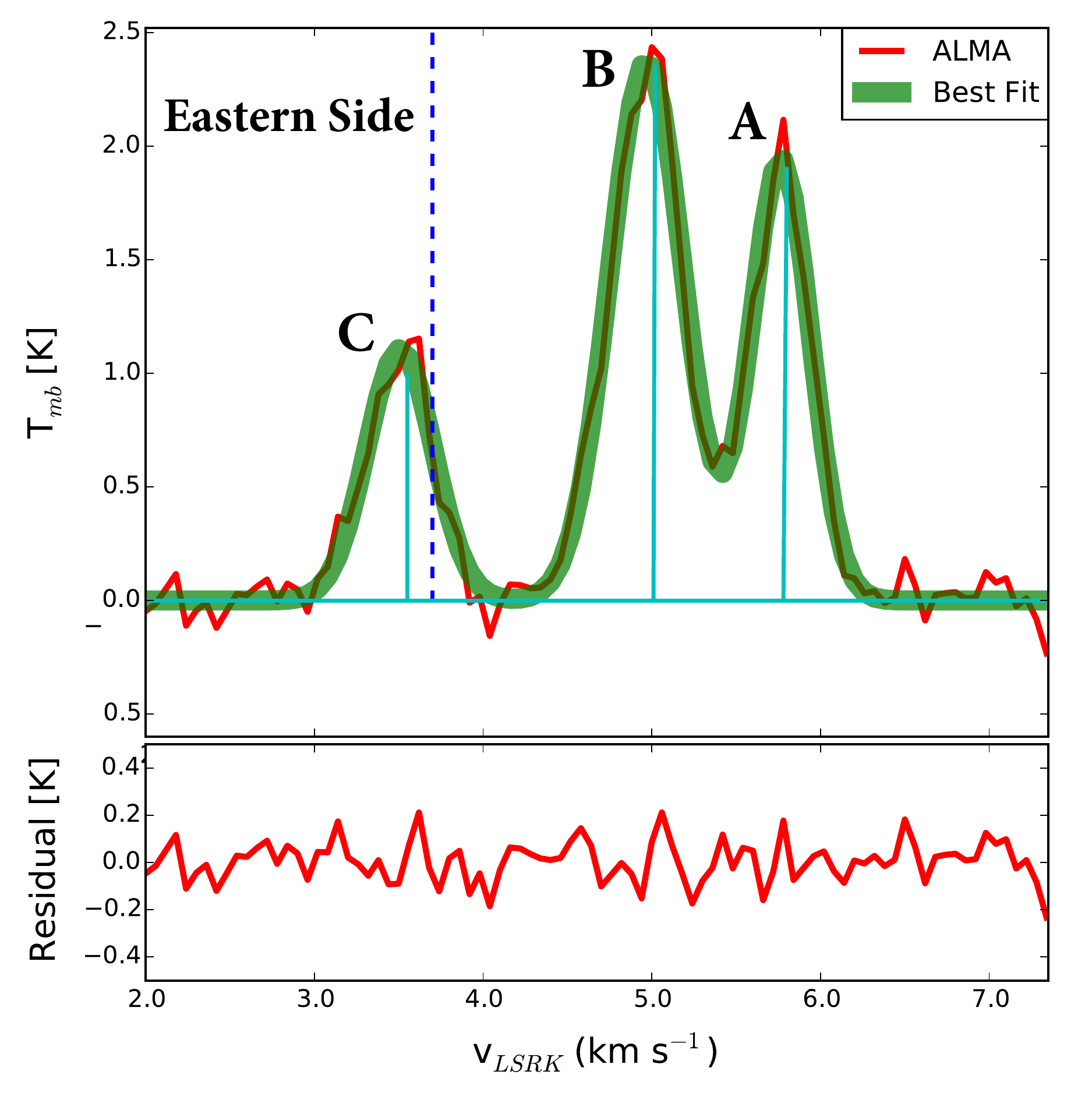}
\caption{The observed CN spectra (red), extracted from the outer region of the disk ($>$ 210 au), with the Gaussian fits (green). Top and bottom panels show the HF structure extracted from the western and eastern sides of the disk, respectively. Cyan lines in the observed spectra mark the brightest HF components and have been scaled to match the peak intensity in the B component of the observed spectrum. The dashed blue line represents the systemic velocity of the system ($\sim$3.7 Km s$^{-1}$). These data can be converted to Jy km s$^{-1}$ by using the unit conversion factor of $\sim$55 K.
\label{Fig:Spectra}}
\end{figure}

The CN HF transitions are well resolved at radii larger than 210 au, on both the east and west sides of the disk. Figure \ref{Fig:Spectra} presents the spectra extracted from ellipse-shaped apertures, centered on regions spatially represented by the cyan ellipses drawn in Fig. \ref{Fig:PV}, for which the HF splitting is well resolved. We used Gaussian fits to these line profiles to determine line widths and relative integrated intensities of the five HF components A--E. The results, for components A to C, are shown in Fig. \ref{Fig:Spectra} for the west and east sides, and results for all 5 components are summarized in Table \ref{Table:Ratios}. Under optically thin local thermodynamic equilibrium (LTE) conditions, the ratio of the intensity of the HF components relative to the brightest component (A/B, C/B, D/B, and E/B) would range between 0.1 and 0.7; the theoretical intensity ratios of the HF components are listed in Table \ref{Table:Ratios}. Under optically thick conditions and for a uniform excitation temperature, the lines would saturate, reaching equal intensities in the limit of very large optical depth.

 
At radii $>$ 210 au, the A, B, C, D and E components are at most marginally optically thick, as determined by the measured HF transition ratios shown in Table \ref{Table:Ratios}. We also note that the estimated HF transition ratios within 210 au (not listed) trend toward 1.0, suggesting the emission is becoming optically thicker. Overall, the D and E components in the CN spectrum are smeared out towards the central source, becoming unresolvable, while the emission in A and C are boosted to intensities nearly comparable to the B component until those components (A+B+C) are blended together into a broad profile centered close to the systemic velocity (3.70 km s$^{-1}$).

\begin{table*}
\centering
\small
\caption{Flux ratios for integrated spectra at radii $>$ 210 au}
  \begin{tabular}{cccccccccc}
  \toprule
&&& \textbf{West$^{b}$} && &&\textbf{East$^{b}$} \\

 \cline{3-5}
 
 \cline{7-9}
 \textbf{HF Component}  &  Intensity Ratio$^{a}$ & Intensity Ratio & FHWM &  $\Delta$V$^{c}$& & Intensity Ratio & FHWM &  $\rm \Delta$V$^{c}$ \\
   & (Theoretical) && & km s$^{-1}$  && &km s$^{-1}$\\
    
     \cline{1-9}
     
A&    0.6311  & 0.68 $\pm$ 0.05 & 0.40 $\pm$ 0.02 & 0.24 $\pm$ 0.01 && 0.54 $\pm$ 0.02& 0.41 $\pm$  0.01& 0.25 $\pm$ 0.01 \\ 
B &    1  & 1  &0.41 $\pm$ 0.01& 0.25 $\pm$ 0.00&  & 1  &0.51 $\pm$  0.01& 0.31  $\pm$  0.01 \\ 
 C&    0.3755  &  0.47 $\pm$ 0.04 &  0.38 $\pm$ 0.02 & 0.23 $\pm$ 0.01 &  & 0.38 $\pm$ 0.02 & 0.47 $\pm$  0.02 & 0.28 $\pm$  0.01\\ 
 D&   0.1196   &  0.15 $\pm$ 0.04 &0.33 $\pm$ 0.06& 0.20  $\pm$ 0.03& & 0.17 $\pm$ 0.02 & 0.42 $\pm$ 0.03 & 0.25 $\pm$ 0.02  \\ 
E &    0.1189  & 0.10 $\pm$ 0.03 &0.30 $\pm$ 0.08 & 0.18 $\pm$ 0.05& & 0.16 $\pm$ 0.02 & 0.40 $\pm$ 0.03 & 0.24 $\pm$ 0.02\\

    \hline
  \end{tabular}
  \label{Table:Ratios}
  
\begin{flushleft}
$^{a}$ Theoretical ratio of intensities of HF transitions to the brightest HF transition (J = 5/2 $\rightarrow$ 3/2, F = 7/2 $\rightarrow$ 5/2).  The relative intensities of HF transitions of CN are adapted from Table 2 of \citep{Punzi2015}.\\
$^{b}$ Values obtained from Gaussian fitting (see Sec. \ref{Sec:Profile}).\\
$^{C}$ $\rm \Delta$V = $\frac{\rm FWHM}{\sqrt{4\: \textup{ln}2}}$.
\end{flushleft}  
 \end{table*}



\subsection{CN Column Densities}
\label{Sec:Column}

The analysis of Sec. \ref{Sec:Profile} provides indications of CN line opacities, but does not address the key physical conditions that give rise to the emission, in particular, CN column densities\footnote{Note that the $\rm N_{CN}$ values derived here represent line-of-sight CN column densities, as opposed to CN column densities measured vertically through the disk.} ($\rm N_{CN}$) and excitation temperatures ($\rm T_{ex}$). To ascertain the range of these and other parameters (e.g., optical depth, $\rm \tau$, and H$_{2}$ number density, $\rm n_{H_{2}}$) as functions of position within the disk, we have modeled the CN HF line structure using a grid-based parameter exploration approach that exploits the non-LTE radiative transfer code RADEX under the so-called Large Velocity Gradient (LVG) framework \citep{vanderTak2007}. RADEX uses an escape probability formalism, with five parameters as inputs:  $\rm n_{H_{2}}$, $\rm N_{CN}$,  kinetic  temperature ($\rm T_{kin}$), background  temperature, and linewidth. To explore a range of thermal parameters consistent with previous studies \citep[e.g.][]{Guilloteau2016,Dutrey2017}, we constructed a 3-D grid consisting of 21 values of H$_{2}$ density in the range $\rm n_{H_{2}}$ =  10$^{4}$ - 10$^{9}$ cm$^{-3}$, 21 values of CN column density in the range $\rm N_{CN}$ = 10$^{11}$ - 10$^{17}$ cm$^{-2}$, and 51 values of $\rm T_{kin}$ from 5 to 60 K. As part of the fitting process (see below), we upsampled these RADEX 3D-grid models by interpolating between grid points to increase the resolution of the parameter space explored. The FWHM line width of all three HF components was held fixed at 0.4 km s$^{-1}$, which is the average of the individual line-widths derived in Sec. \ref{Sec:Profile}. For the background temperature, we adopt 2.73 K.  For CN emission, RADEX then predicts $\rm T_{ex}$ and $\rm \tau$ values for the `A', `B', and  `C'  components from each set of these input parameters ($\rm T_{kin}$, $\rm n_{H_{2}}$, linewidth, $\rm N_{CN}$) using collision rates from \citet{HilyBlant2013}. Since RADEX does not treat line overlaps, we restricted the spectral modeling to disk regions where these CN HF structures are well resolved (see below).

To perform HF line fitting covering the full spectral range of the `A' $+$  `B' $+$  `C'  components, we use a forward-fitting approach built into the PySpecKit Python package \citep{Ginsburg2011}. For each grid point (i.e., combination of $\rm T_{kin}$, $\rm n_{H_{2}}$, and $\rm N_{CN}$), the RADEX-computed $\rm T_{ex}$ and $\rm \tau$ parameters are used to generate synthetic line profiles for each of the three HF components from
 \begin{equation}
\rm T_{MB}(\nu ) = \rm T_{o}\left ( \frac{1}{e^{\frac{T_{o}}{T_{ex}}}-1} - \frac{1}{e^{\frac{T_{o}}{2.73}}-1}   \right )\left ( 1 - e^{-\tau } \right ), 
\label{Eq:RT}
\end{equation}
where $ \rm T_{MB}(\nu)$ is the predicted brightness temperature, $ \rm T_{o}(\nu ) = \frac{h\nu }{k}$,  $k$ is Boltzmann's constant, and $h$ is Planck's  constant \citep[e.g.][]{Mangum2015}. The complete synthetic spectrum of the CN 2--1 lines for a given parameter set is obtained as the sum of intensities of the individual HF components, each described by Eq. \ref{Eq:RT}. This procedure is validated a posteriori because the excitation temperatures of each component are very similar and line overlap for the A and B components is modest (see Figures \ref{Fig:Spectra} and \ref{Fig:Radex}). The set of synthetic spectra is then compared to the observed spectra using a $\chi^{2}$ minimization process implemented in the PySpecKit Python package,  which employs the Levenberg-Marquardt optimization method via the MPFIT algorithm \citep[][ and references within]{Markwardt2009}. 

\begin{figure}
\centering
\includegraphics[width=0.49\textwidth]{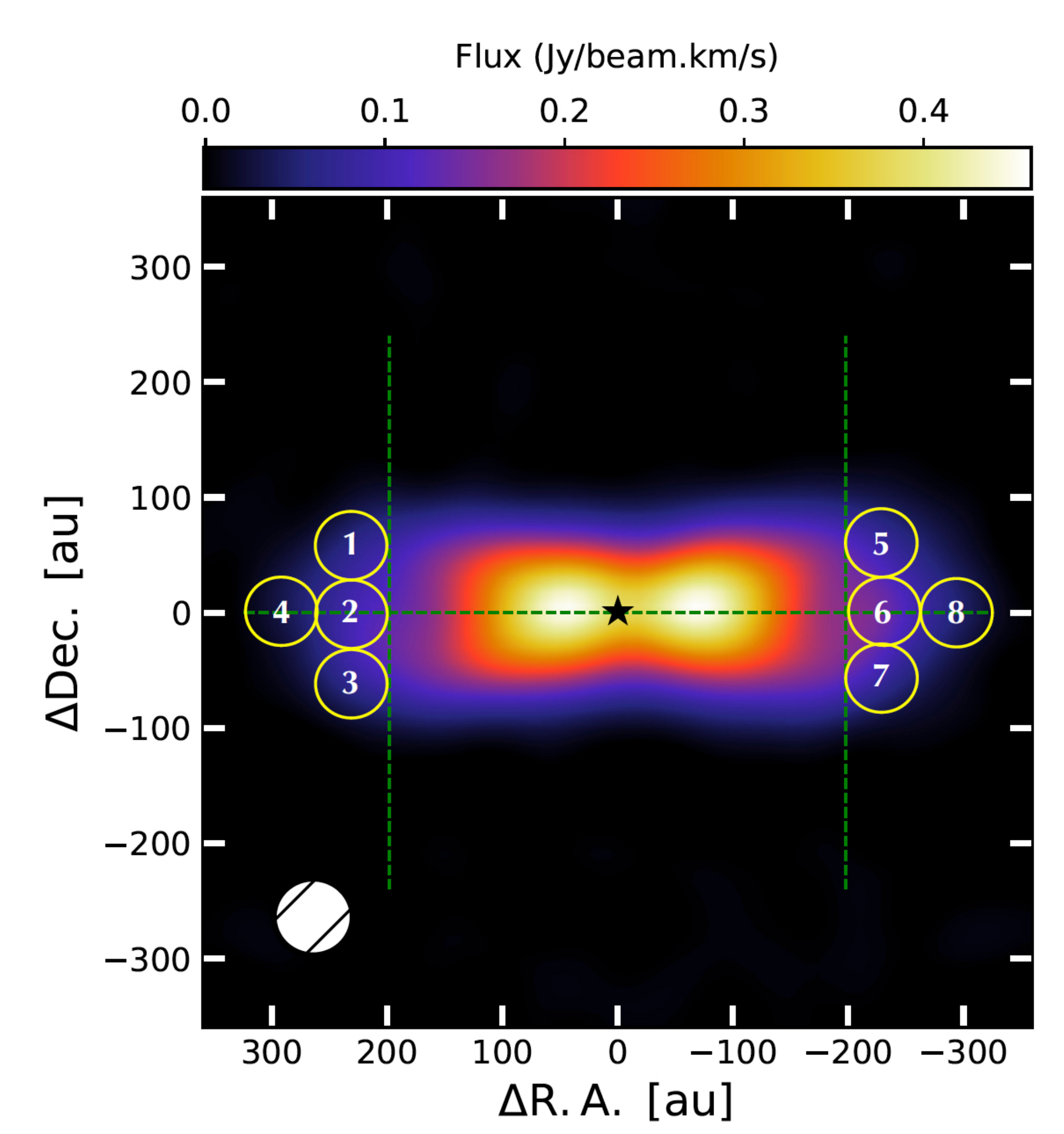}
\caption{Integrated intensity map of the CN emission overlaid with eight enumerated circles representing the eight single-beamwidth (0.3$^{''}$ radius) regions from which we extract spectra at each side of the disk at radii $>$ 210 au (see Table~\ref{Table:Column}, Section \ref{Sec:Column}, and Fig. \ref{Fig:Radex}). The synthesized beam size is indicated by the white ellipse. Vertical green dashed lines are placed at radii of $\sim$210 au from the central star; the horizontal green dashed line indicates the disk midplane..
\label{Fig:CN_regions}}
\end{figure}

\begin{figure*}
\centering

\includegraphics[width=0.31\textwidth,]{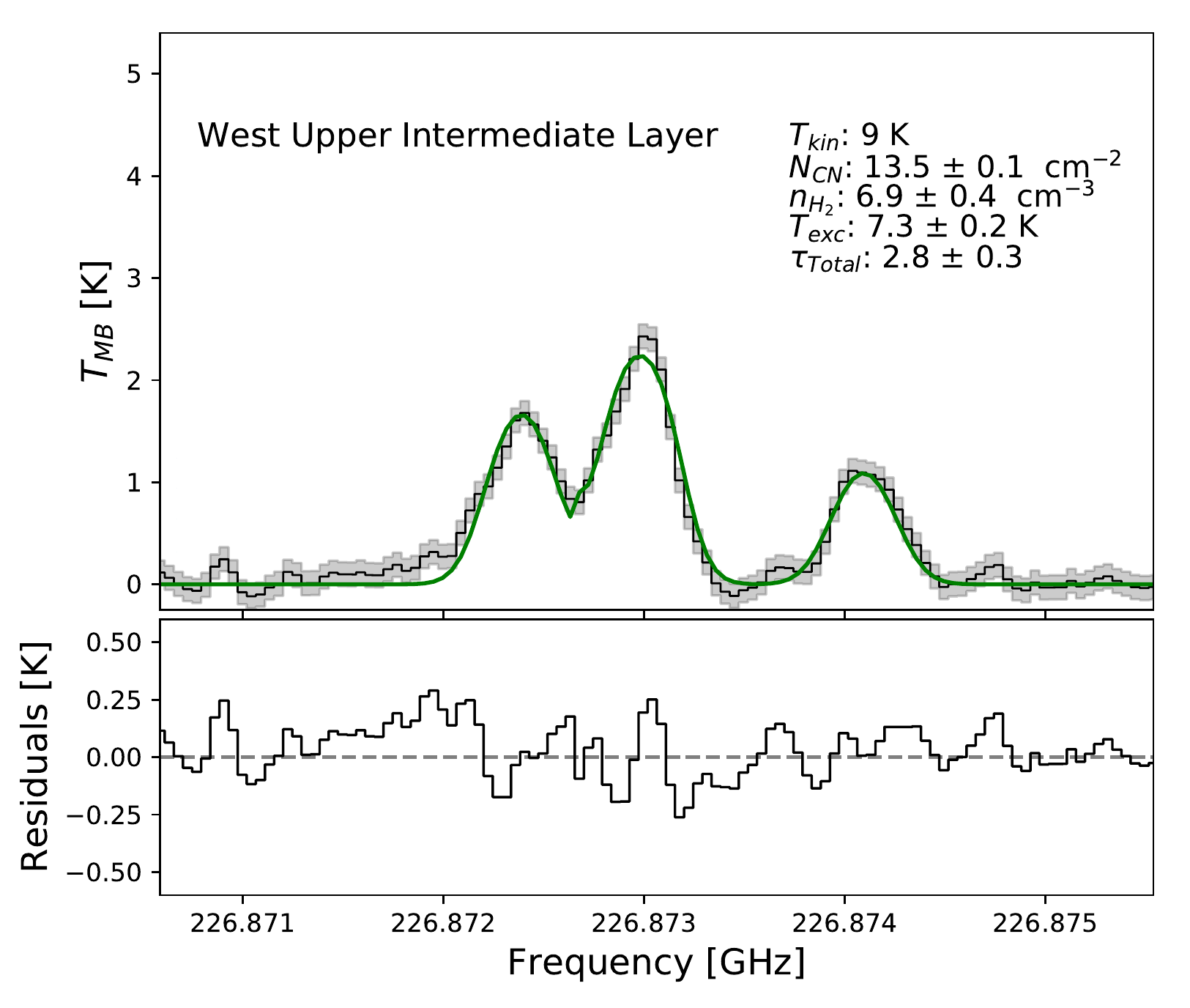}
\includegraphics[width=0.31\textwidth]{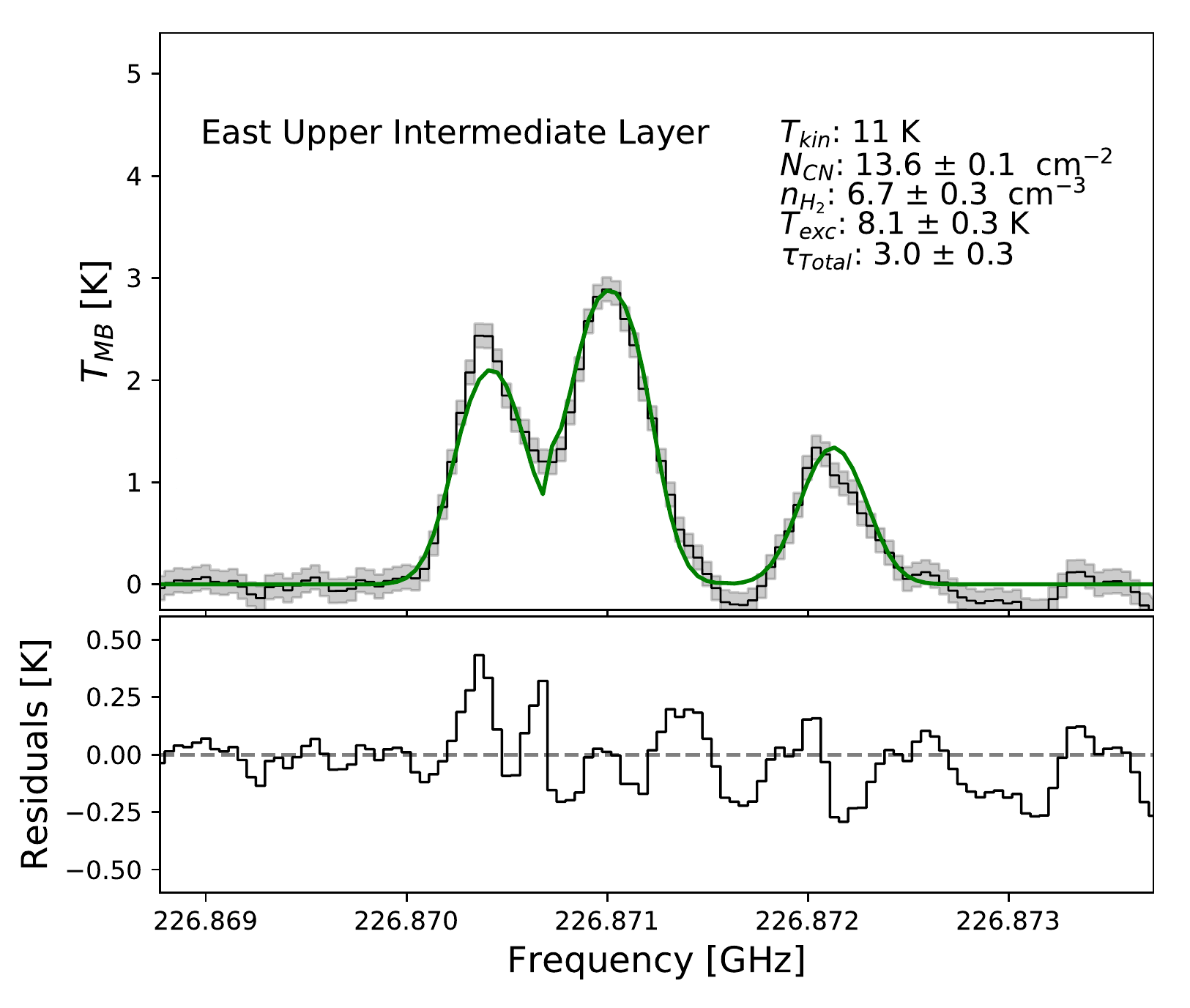}

\includegraphics[width=0.31\textwidth]{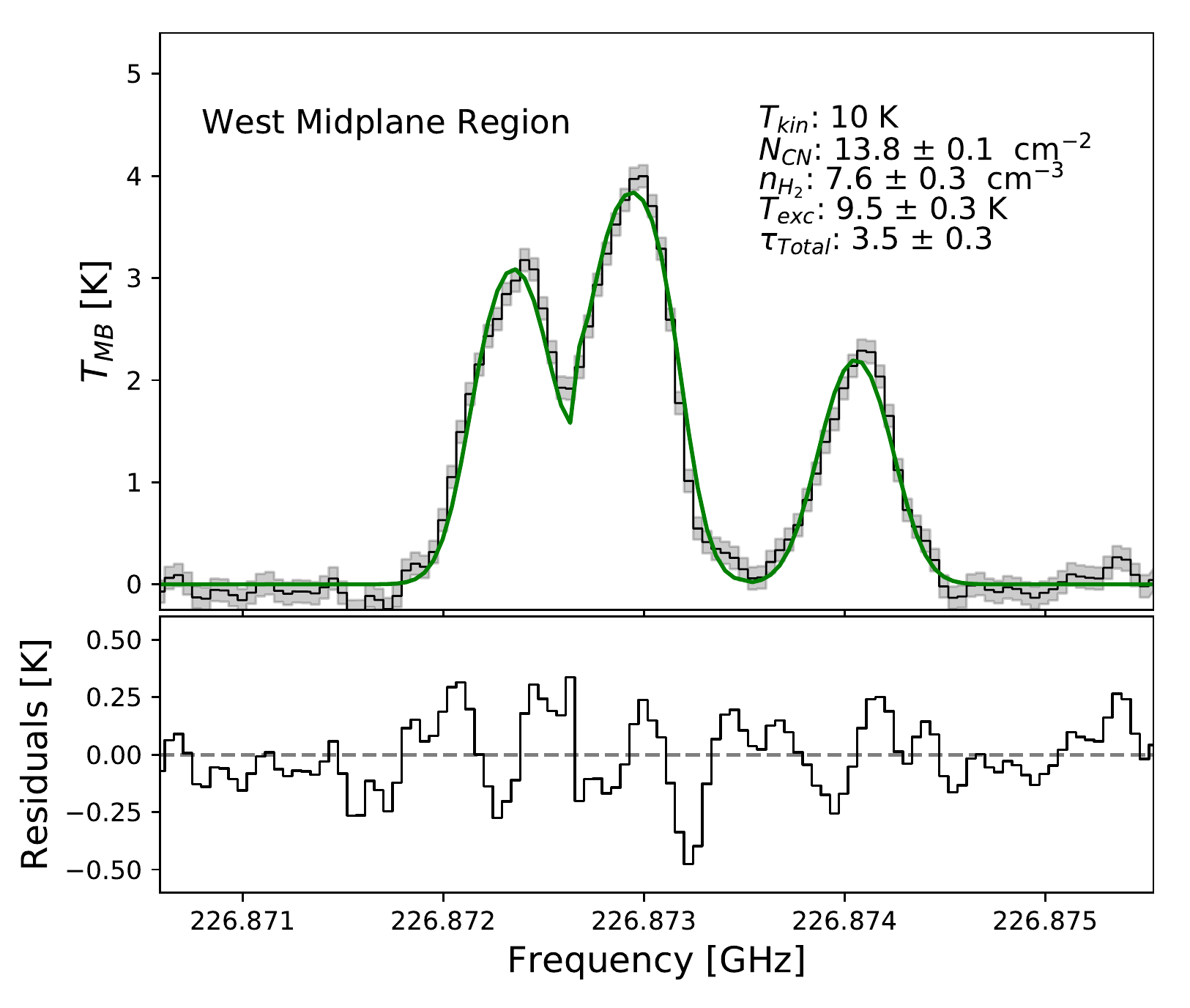}
\includegraphics[width=0.31\textwidth]{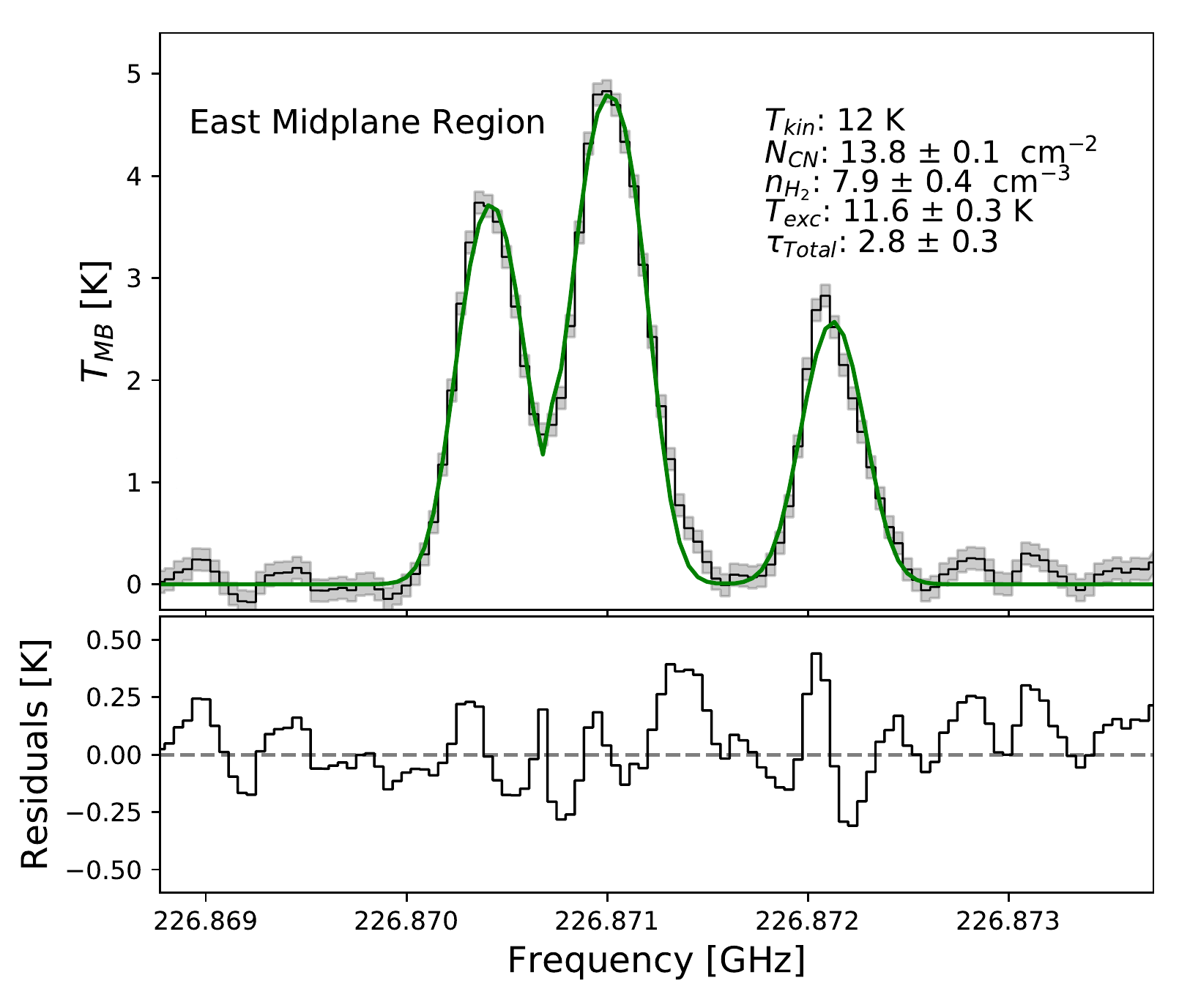}

\includegraphics[width=0.31\textwidth]{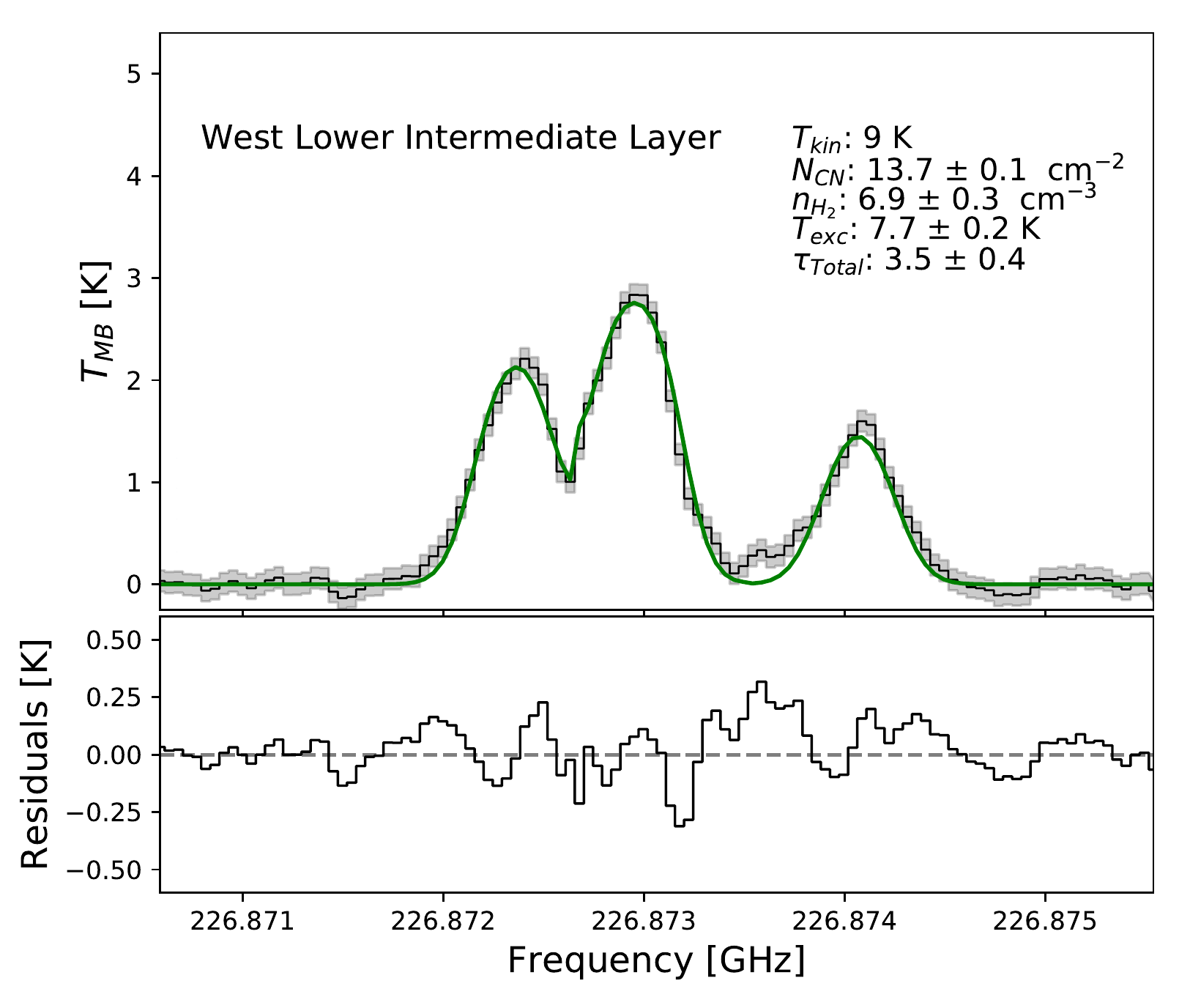}
\includegraphics[width=0.31\textwidth]{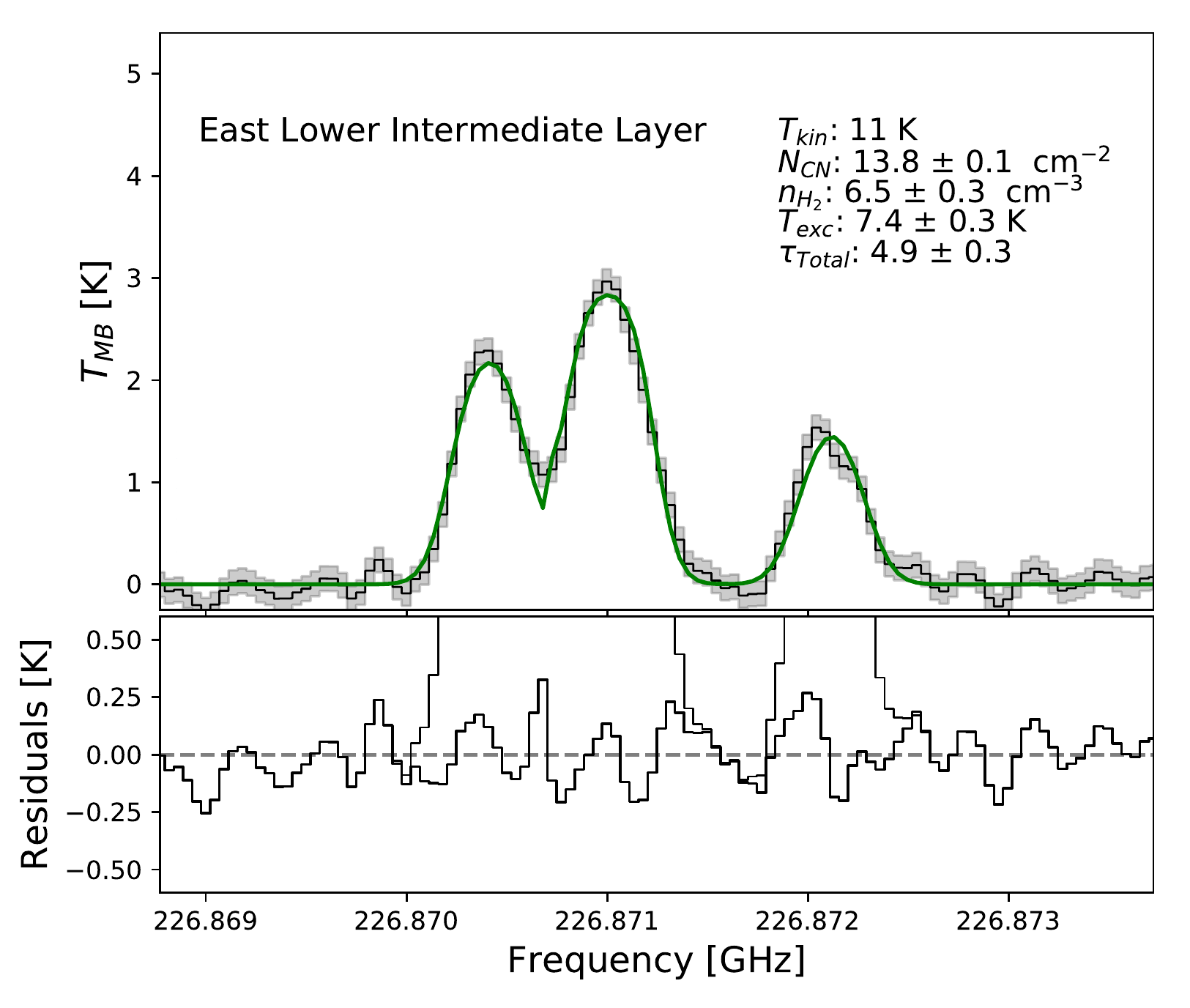}

\includegraphics[width=0.31\textwidth]{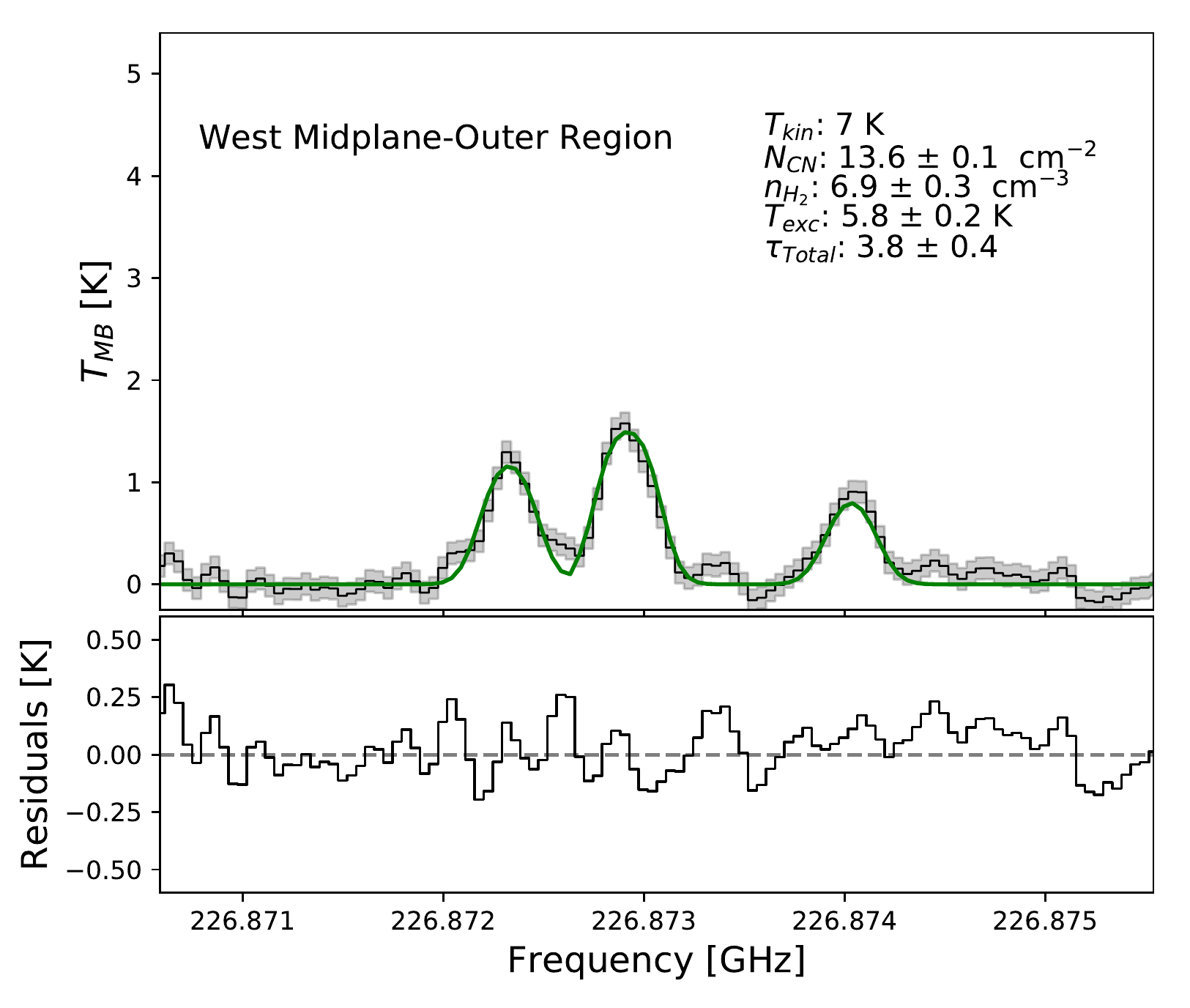}
\includegraphics[width=0.31\textwidth]{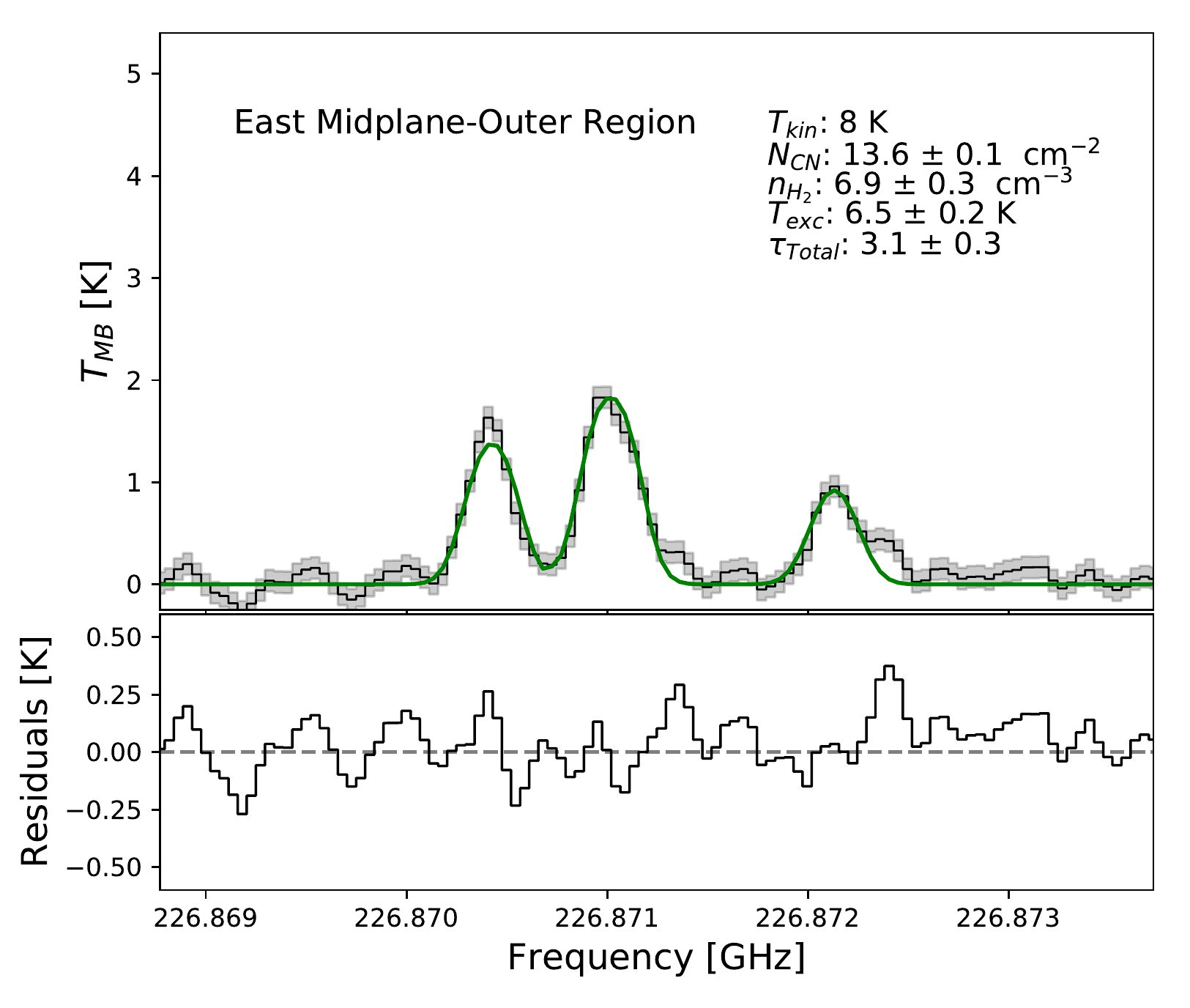}

\caption{Spectra of the integrated CN (2-1) emission from the upper and lower surfaces, midplane and outer midplane of the disk along with the simultaneous fits to components `A',  `B' , and  `C'; see Table \ref{Table:Ratios}. The best-fit LVG models from the $\chi^{2}$ minimization are displayed as a green line. The fitted parameters are the total optical depth ($\rm \tau$; `A' +  `B' +  `C'), excitation temperature ($\rm T_{ex}$), and on logarithmic scale, CN column density ($\rm N_{CN}$) and particle density ($\rm n_{H_{2}}$); see Table \ref{Table:Column}. 
Spectra in the left and right columns are extracted from the west and east sides of the disk, respectively. The fit residuals are shown as a function of frequency at the bottom of each panel. The shaded gray area shows the error computed from a signal-free part of the spectra.
\label{Fig:Radex}}
\end{figure*}

The preceding analysis was carried out for eight specific outer disk regions (r $>$ 210 au) where the HF components are spectrally well resolved, on both the east and west sides of the disk; see Figure \ref{Fig:CN_regions} and Table \ref{Table:Column}.
Figure \ref{Fig:Radex} displays the CN spectra extracted from these eight regions --- each of diameter one beamwidth ($\sim$0.6$^{''}$), centered at upper and lower intermediate layers, midplane, and outer midplane, on both the east and west sides of the disk (Figure \ref{Fig:CN_regions}) --- overlaid with the best-fit LVG models, i.e., the model with the lowest $\chi^{2}$ value. The values of $\rm N_{CN}$, $\rm n_{H_{2}}$, $\rm T_{kin}$, $\rm T_{ex}$, and $\rm \tau$ obtained for each region from this $\chi^{2}$ minimization process are listed in Table \ref{Table:Column}. We caution that --- as is typical in such modeling --- there is an inherent degeneracy, wherein the HF lines are well reproduced either by very dense and cold gas or by lower density, warmer gas.


These results (Table \ref{Table:Column}) indicate that the gas density traced by CN in the regions we modeled (Figure \ref{Fig:CN_regions}) is a few $\times10^{6}$ cm$^{-3}$, and that the gas is quite cold, with a best-fit $\rm T_{kin}$ in the range 10--12 K. We consistently find $\rm T_{ex}$ $<$  $\rm T_{kin}$, where $\rm T_{ex}$/$\rm T_{kin}$ decreases as $\rm T_{kin}$ increases, suggesting the gas may be somewhat subthermally excited. For $\rm T_{kin}$ $>$ 25 K, we find the models fail to reproduce (underestimate) the intensity of the C component of the CN HF structure.
The best-fit CN column densities are consistently $\sim$10$^{13}$ cm$^{-2}$. The modeling results also indicate that the west side of the disk may be somewhat colder. We caution, however, that the results for $\rm T_{kin}$ are subject to systematic uncertainties (arising from, e.g., the various model assumptions and the specific approach to $\chi^{2}$ minimization) that are difficult to quantify. The possibility of such an east-west asymmetry in the disk warrants followup via CN mapping at higher resolution.

\begin{table*}
\centering
\caption{RADEX Model Parameters.
}
  \begin{tabular}{ccccccccc}
  \toprule
    
No. & Region &  \multicolumn{2}{c}{\textbf{Location}}  & $\rm T_{kin}$ & $\rm n_{H_{2}}$&  $\rm N_{CN}$ & $\rm T_{ex}$ & $\rm \tau$$^{a}$   \\
   
& & \textbf{$\Delta$ R.A.} &  \textbf{$\Delta$ Dec}. &  [K] &$\times$ 10$^{6}$ [cm$^{-3}$] &$\times$ 10$^{13}$ [cm$^{-2}$]  & [K] & \\
&&[au]&[au]&&\\

     \cline{1-9}
     
1. &W. Upper Intermediate Layer & \textbf{$+$ 230}& \textbf{$+$ 50}  &  9 & 6.8& 3.5 &  7.3 & 2.8   \\ 
2. & W. Midplane & \textbf{$+$ 230}&  \textbf{$+$ 0.0}& 10 & 50.0 &  6.0  & 9.5 & 3.5     \\ 
3. & W. Lower Intermediate Layer & \textbf{$+$ 230}&  \textbf{$-$ 50} & 9  & 8.0 & 4.4& 7.7 & 3.5  \\ 
4. &W. Outer Midplane  & \textbf{$+$ 280}&  \textbf{$+$ 0.0} & 7  & 7.3   & 4.2 & 5.8 & 3.8   \\ 

   \cline{1-9}

5. & E. Upper Intermediate Layer & \textbf{$-$ 230}&  \textbf{$+$ 50}  & 11 & 5.4 & 3.9 & 8.1 & 3.0   \\ 
6. & E. Midplane & \textbf{$-$ 230} &  \textbf{$+$ 0.0}& 12& 74.0 & 5.7   &  11.6 & 2.8 \\ 
7. & E. Lower Intermediate Layer & \textbf{$-$ 230}&  \textbf{$-$ 50} & 11   & 3.2 & 6.4 & 7.4 & 4.9  \\ 
8. & E. Outer Midplane &\textbf{$-$ 280}&  \textbf{$+$ 0.0} & 8  &  7.0 &  3.7 & 6.5 & 3.1  \\

    \hline
  \end{tabular}
  \label{Table:Column}
  
\begin{flushleft}
$^{a}$ Sum of the opacities at line center of A, B and C HF components.\\
\end{flushleft}  
 \end{table*}

\section{Discussion}
\label{Sec:Discussion}

\begin{figure}
\centering
\includegraphics[width=0.49\textwidth]{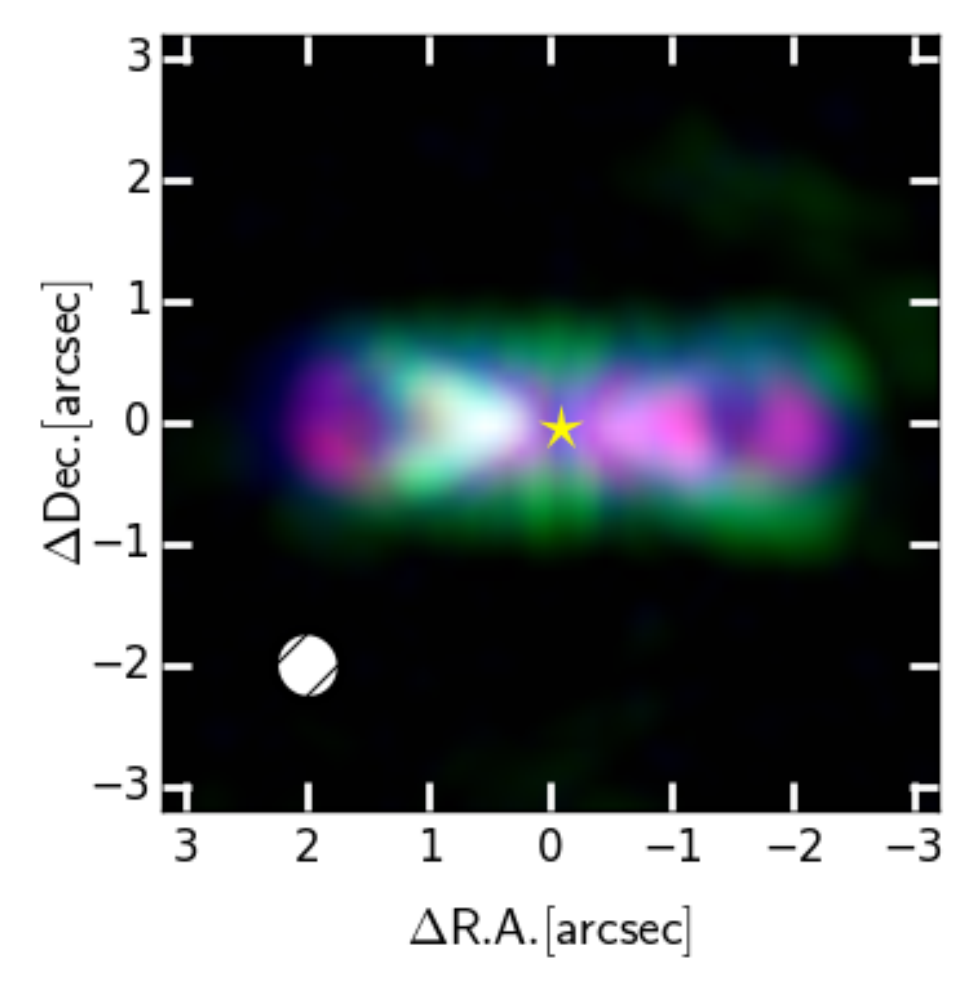}
\includegraphics[width=0.49\textwidth]{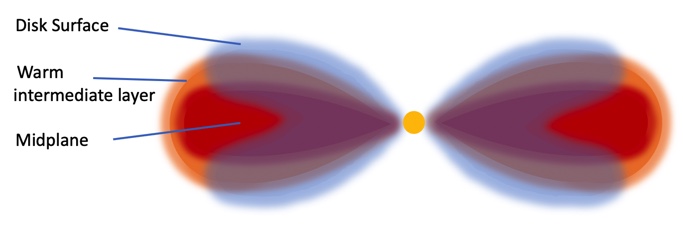}
\caption{Top: Three-color (RGB) image generated from CS, CN, and CO peak intensity (CASA moment 8) maps. CS, CN, and CO are assigned to the red, green, and blue channels, respectively; each color channel is scaled linearly. Bottom: Schematic illustration of the different layers traced by CS (midplane), CN (midplane to intermediate layers) and CO (upper disk surface) emission.  
\label{Fig:Cartoon}}
\end{figure}

\subsection{Vertical Molecular Emission Structure}
\label{Sec:Gradient}

The analysis presented in Sec.~\ref{Sec:Scale} (see, in particular Fig.~\ref{Fig:layers_radii}), establishes that the CS, CN, and CO emission regions in the Flying Saucer disk are vertically stratified. To further demonstrate this vertical stratification, we have generated maps of the peak intensities of CN, CS, and CO (i.e., CASA ``moment 8'' maps) from their respective image cubes (presented in Figs.~\ref{Fig:Channel}, \ref{Fig:CS_Channel}, and \ref{Fig:CO_Channel}). These peak emission images are rendered as an RGB image in Fig. \ref{Fig:Cartoon}a, where they are compared with a schematic illustration of the disk structure in these three tracers (Fig.~\ref{Fig:Cartoon}b).

The structure illustrated in Fig. \ref{Fig:Cartoon} can be summarized as follows: CO reaches the largest vertical extension, delineating the disk surface, with a deficit of CO near the midplane; CS emission is confined to the layers closest to the cold disk midplane; and CN traces layers intermediate between these extremes, albeit with some overlap with CS, in that the CN also appears to extend to near the midplane. As discussed in \citet{Dutrey2017}, the midplane deficit of CO seems to correspond to the vertical settling of dust grains toward the midplane as traced by the 242 GHz continuum emission (Fig. \ref{Fig:cuts}d), which forms an elongated structure of about $\sim$190 au in radius \citep[see also][]{Guilloteau2016}. Of the three molecular tracers in the ALMA data thus far available for the Flying Saucer, it is evident from Fig.~\ref{Fig:layers_radii} and Fig. \ref{Fig:Cartoon}a that CS most closely traces this (vertically and radially confined) continuum emission, although CS is clearly more extended (in both dimensions). The CN originates from a layer that is radially more extended than and elevated with respect to CS. In addition, there appears to be a dearth of CN emission at the midplane at radii between $\sim$80 and 200 au; this is apparent in the CN velocity channel maps (Fig. \ref{Fig:Channel}). At the outer regions of the disk, the CN emission arises from the upper and lower intermediate layers and extends down to the midplane, outlining the disk radial extension. Possible mechanisms that might enhance CN production in these (intermediate) disk layers are discussed in  Sec.~\ref{Sec:Location}.



\citet{Teague2016} previously suggested such a vertically stratified emission structure for CS, CN, and CO emission lines in the case of the nearly face-on TW Hydrae disk (TW Hya). From analysis of multiple transitions of the three species, these investigators derived CO, CN, and CS excitation temperatures of $\sim$35, 25, and 12 K, respectively \citep[see also][]{Teague2020}. On this basis, \citet{Teague2016} deduced that the emission from the three molecules arises from decreasing heights above the disk midplane. This conclusion is consistent with the data presently available for the nearly edge-on Flying Saucer (Fig. \ref{Fig:Cartoon}), although we deduce colder excitation (and kinetic) temperatures for CN in this disk (see Sec.~\ref{Sec:Location}). However, the implied discrepancy in the thermal structures of the TW Hya and Flying Saucer disks is perhaps not surprising, given the different stellar parameters and ages of both systems. In particular, the difference in the derived values of $\rm T_{kin}$ might lie with the fact that the TW Hya is a late-K star of rather advanced age \citep[$\sim$8 Myr; e.g.][]{Donaldson2016}, and is a luminous source of high-energy photons, with L$_{X}$ = 1.4 $\times$ 10$^{30}$ ergs s$^{-1}$ \citep{Kastner2002}, whereas the Flying Saucer central star is a younger, M-type protostar that has thus far gone undetected in X-rays \footnote{Chandra and XMM-Newton X-ray observations of a $\rho$ Oph field that included Flying Saucer were obtained in 2000 and 2003 with exposure times of $\sim$ 5 ks and $\sim$ 34 ks, respectively. Neither observation yield a detection at the position of the object. Given the disk's edge-on orientation, however, it is difficult to draw conclusions as to the intrinsic X-ray luminosity of the central star.}. Furthermore, the TW Hya disk has a CN emission extension of $\sim$180 au \citep{Vlemmings2019, Teague2020}, a factor of $\sim$2 smaller than that of the Flying Saucer disk.

\subsection{CN Physical Conditions}
\label{Sec:Temp}

The results of our RADEX analysis indicate that the CN traces cold gas, and is optically thick, in the disk regions modeled via RADEX, i.e., $\rm T_{kin}$ $\sim$ 7-10 K (with total $\rm \tau$ $\sim$ 3-4) and $\rm T_{kin}$ $\sim$ 8-12 K (with total $\rm \tau$ $\sim$ 3-5) at radii $>$210 au and heights $<$60 au in the western and eastern sides of the disk, respectively (Sec \ref{Sec:Column}). The low midplane temperatures appear compatible with a model of the temperature structure of the Flying Saucer disk obtained by \citet{Dutrey2017} from numerical simulations of the disk's CO emission. Specifically, those authors found $T\sim 5-10$ K in the disk midplane at offsets of $\sim$230-280 au \citep[][their Fig.~8g]{Dutrey2017}, corresponding to the western and eastern midplane regions we included in our analysis (Table \ref{Table:Column}). Furthermore, the densities we obtain from our line profile modeling ($\sim$$10^{6}$ to $10^{7}$ cm$^{-3}$; Table 4) are generally consistent with, if somewhat smaller than, the densities predicted for these same (external) disk regions in the Dutrey et al. (2017) modeling (their Fig. 7). Their modeling also indicates, however, that $T_{kin} \sim$ 15--25 K at heights of $\sim$50 au at the same ($\sim$230-280 au) radial offsets. These disk regions correspond to the ``intermediate layers'' for which we find $\rm T_{kin}$ $\sim$ 9--11 K based on RADEX line profile fitting (Table \ref{Table:Column}), and are not significantly different from the corresponding values in the midplane. This lack of a significant vertical gradient in $\rm T_{kin}$ at radii $>$210 au, as inferred from CN emission, may indicate that the standard protoplanetary disk model, on which the \citet{Dutrey2017} modeling is based, may not well describe the vertical temperature structure in the Flying Saucer.

On the other hand, our RADEX modeling also yields evidence that the CN molecules are subthermally excited ($T_{ex} < T_{kin}$) away from the midplane, with the divergence between $\rm T_{ex}$ and $\rm T_{kin}$ more evident on the eastern side of the disk  (Table \ref{Table:Column}). We caution that, as noted earlier, it remains to be determined whether the apparent differences in the CN thermalization inferred for the two sides of the disk reflect a real asymmetry in disk physical conditions. Indeed, one difficulty with the low $\rm T_{kin}$ inferred here is that one would expect CN and CS to be severely depleted from the gas phase in disk regions in which $\rm T_{kin}$ drops well below the freeze-out temperature of CO (i.e., $\sim$17 - 19 K). However, subthermal excitation becomes more likely in disk regions with a significant degree of depletion of dust grains, as may characterize the Flying Saucer at radii $>210$ au. Previous studies have shown that when the gas-to-dust ratio drops to $\sim$20 (or $n_{H_2}$ drops by a factor $\sim$5), CN would be produced under subthermal excitation conditions, although this is only expected in regions where the density is of the order of a few 10$^{5}$ cm$^{-3}$ \citep{Chapillon2012}.

\subsection{CN: Location and Formation}
\label{Sec:Location}

\citet{Cazzoletti2018} modeled CN emission from protoplanetary disks under the assumption that CN formation mainly proceeds through far ultraviolet (FUV) pumping of H$_{2}$ to vibrationally excited levels, in addition to photodissociation of HCN. As a main result, the highest CN abundance should be located in the upper region of the disk, while in outer disk regions, the emission extends down to the midplane, forming a CN emission ring. Although the HF transitions are not considered specifically in this model, the CN features observed in the Flying Saucer  disk appear reasonably consistent with the predicted formation pathway, where CN is produced near the surface and outer regions of the disk (Fig. \ref{Fig:Channel}). However, it is apparent from Fig.  \ref{Fig:layers_radii} that the CN layer lies between the CO and CS layers, clearly showing that CN emission traces intermediate layers rather than disk surface layers. In addition, the strong link between UV radiation and CN formation would appear to require that most of the CN emission should arise from upper disk layers, given that the FUV radiation field is attenuated by dust at moderate disk depths \citep{Gorti2009}. Indeed, the notion that UV controls the abundance of CN is consistent with the deficit of midplane CN emission at radii between $\sim$80 and 200 au (see Fig. \ref{Fig:Channel}), matching the spatial extension of the 242 GHz continuum. 

However, as noted earlier, we also find that the CN arises from very cold ($\rm T_{kin}$  $\lesssim$15 K) regions of the disk (Sec.~\ref{Sec:Temp}). This suggests that FUV radiation alone may be insufficient to explain the observed CN abundances, and one must invoke more deeply penetrating sources of high energy radiation, such as X-rays \citep[see discussions in][]{Kastner2014, HilyBlant2017}. Indeed, the bulk of the CN intensity is localized within the central $\sim$0.8$^{''}$ ($\sim$ 100 au) in radius (Fig. \ref{Fig:cuts}b) of the disk, a region where enhancement of CN production can be produced by exposing a disk that is in advanced stages of settling to X-rays or a combination of X-rays and FUV radiation \citep{Stauber2005, Stauber2007}.
Moreover, recent simulations from detailed physical/chemical disk models indicate that the distribution of radicals such as CN could be present as multiple-layered structures \citep[e.g.][]{Ruaud2019}. In fact, several channel maps of the CN line, Fig. \ref{Fig:Channel}, appear to show emission arising from the midplane, especially at the inner part of the CN emission `ring', while there are velocities in the channel maps at which it appears that CN arises from intermediate layers, at larger disk radii. These structures are qualitatively similar to the predictions of the \citet[][]{Ruaud2019} models. Both emission regions may contribute to low-temperature CN emission, such as we infer from our RADEX modeling (Sec. \ref{Sec:Column}). Higher-resolution ALMA imaging spectroscopy of multiple transitions of CN would help establish whether there are indeed two distinct disk layers involved in efficient CN production.

\section{Summary}

Using archival ALMA data, we have investigated the vertical chemical and physical structure of the edge-on Flying Saucer disk. This is in order to constrain CN formation pathways that took place during disk evolution and planetary formation. In particular, we used RADEX models to generate CN (2-1) synthetic spectra, and thus predict the disk thermal conditions that characterize the observed CN emission from the Flying Saucer  disk. As a comparative approach, we have used molecular lines tracing different disk physical conditions, specifically$^{12}$CO and CS, in order to further constrain spatially the location of the CN emission. Our main results are as follows:

\begin{itemize}

\item We demonstrate that $^{12}$CO, CN, and CS molecular lines can be used to trace the vertical structure of edge-on disks. We estimated the scale heights of these molecules, and they reveal a stratified structure within the flared Flying Saucer  disk. Basically,  $^{12}$CO traces the low-density disk surface, while CN and CS trace the denser intermediate layers and midplane region, respectively. 

\item The analysis of the CN channel maps reveals complex structures superimposed on Keplerian gas kinematics of the Flying Saucer  disk. An outstanding feature is the CN emission deficit near the midplane at radii $>$ 100 au, predominantly arising from intermediate layers. However, in the inner regions of disk, CN emission appears to arise from layers closer to the midplane. This variation of the CN vertical location along the disk might be related to the formation route of CN, whose production is likely related to the degree of irradiation of the flared disk by X-rays and UV photons from the central star.

\item Based on RADEX non-LTE modeling of the three brightest HF components of the CN(2-1) line, we estimated thermal parameters in different regions of the disk. Our results characterized a disk with thermal emission from the midplane and sub-thermal conditions in intermediate layers. Near the disk midplane, where we derive densities $\rm n_{H_{2}}$ $\sim$10$^{7}$ cm$^{-3}$ at relatively low $\rm T_{kin}$ ($\sim$ 12 K), we find that CN is thermalized, while sub-thermal, non-LTE conditions appear to obtain for CN emission from higher (intermediate) disk layers.

\item Higher-resolution ALMA imaging spectroscopy of multiple transitions of CN is required to confirm the cold outer disk temperatures we infer from our radiative transfer modeling. Such followup observations would also help establish whether there exist multiple, distinct disk layers involved in efficient CN production.

\end{itemize}

\begin{acknowledgements}
We wish to thank the anonymous referee for providing insightful comments.

This  paper  makes  use  of  the  following  ALMA  data: ADS/JAO.ALMA$\#$2013.1.00387.S.  ALMA  is  a  partnership of  ESO (representing its member states), NSF  (USA)  and  NINS  (Japan),  together  with  NRC (Canada),  MOST  and  ASIAA  (Taiwan),  and  KASI (Republic of Korea), in cooperation with the Republic of  Chile. The Joint ALMA Observatory is operated by ESO, AUI/NRAO and NAOJ. The National Radio Astronomy Observatory is a facility of the National Science Foundation operated under cooperative agreement by Associated Universities, Inc.

\end{acknowledgements}

\bibliography{refs}
\bibliographystyle{aa}

\begin{appendix}

\section{CN HF Components}
\label{App:A}

\begin{figure}
\centering
\includegraphics[width=0.49\textwidth]{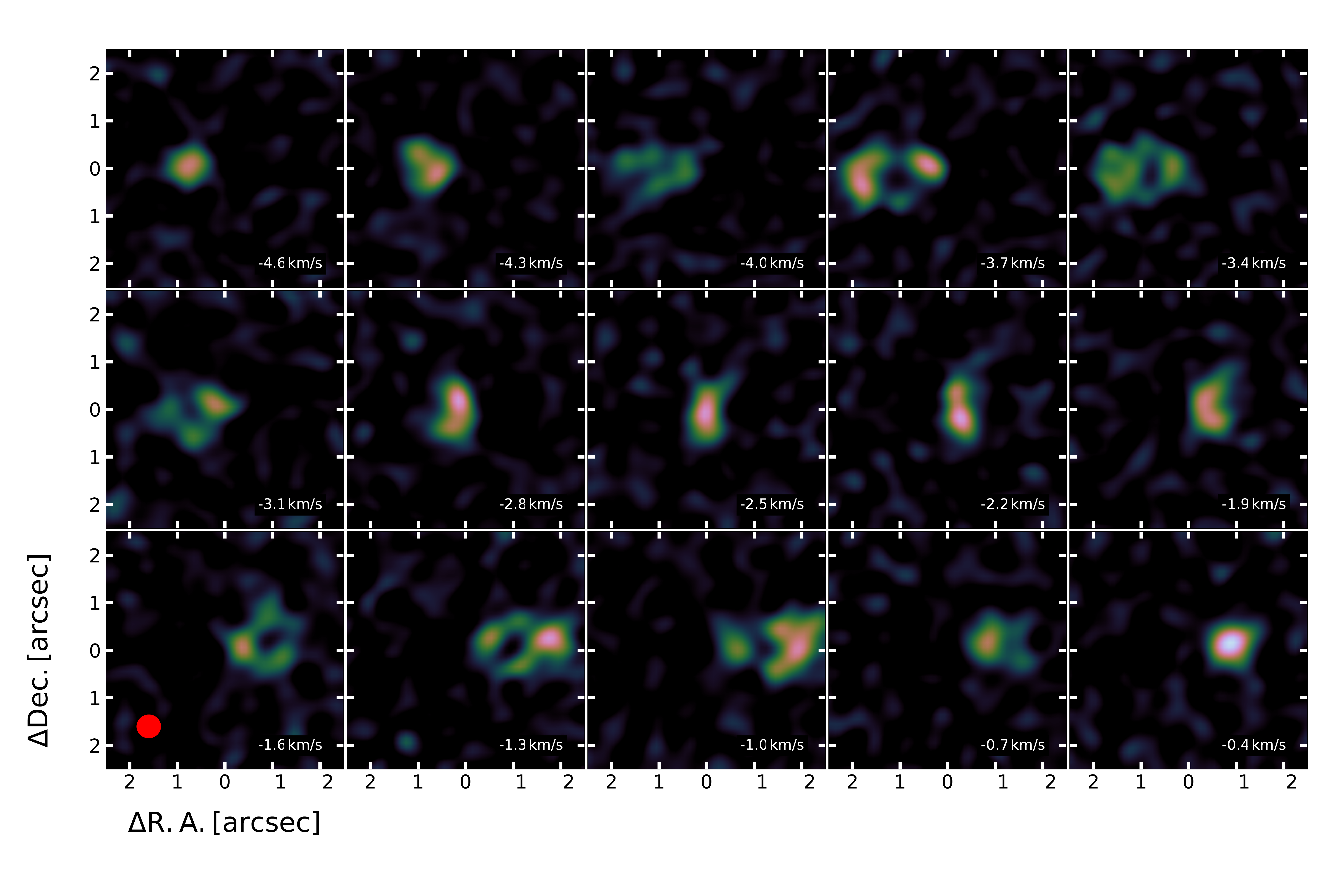}
\includegraphics[width=0.49\textwidth]{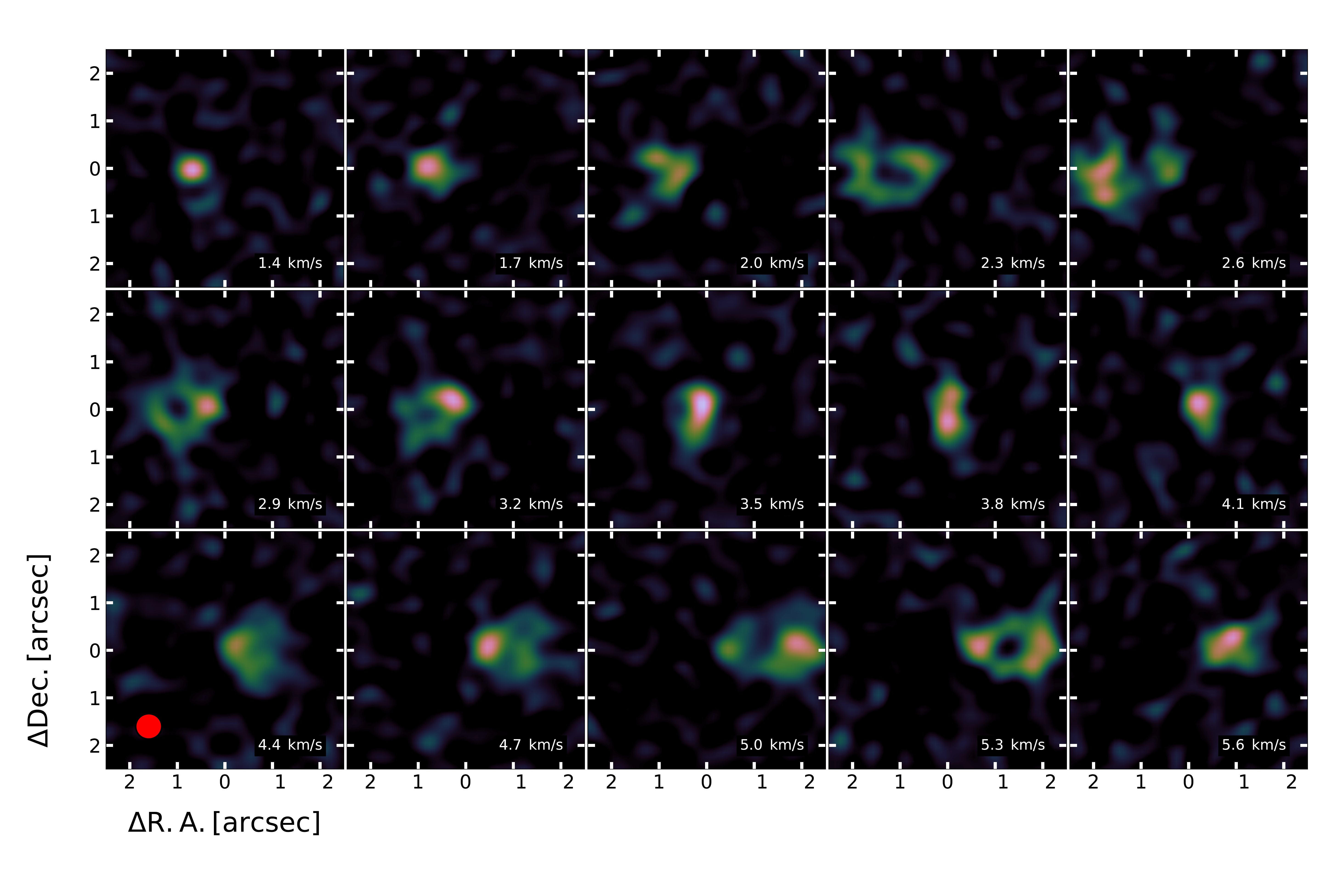}
\caption{Channel maps of the CN emission in the Flying Saucer from -4.60 to -0.40 km $^{-1}$, and from 1.40 to 5.60 km $^{-1}$.  The frequency range is presented for the `D' (Top), and `E'  (Bottom) HF components, which are J = 5/2 $\rightarrow$ 3/2, F = 3/2 $\rightarrow$ 3/2,  and J = 5/2 $\rightarrow$ 3/2, F = 5/2 $\rightarrow$ 5/2, respectively (see Table \ref{Table:Ratios}). The spectral resolution is 0.3 km s$^{-1}$ and the rms per channel is 2.9 mJy beam$^{-1}$. The synthesized beam size is represented in the lower left panel. 
\label{Fig:Channel_DE}}
\end{figure}

\begin{figure}
\centering
\includegraphics[width=0.49\textwidth]{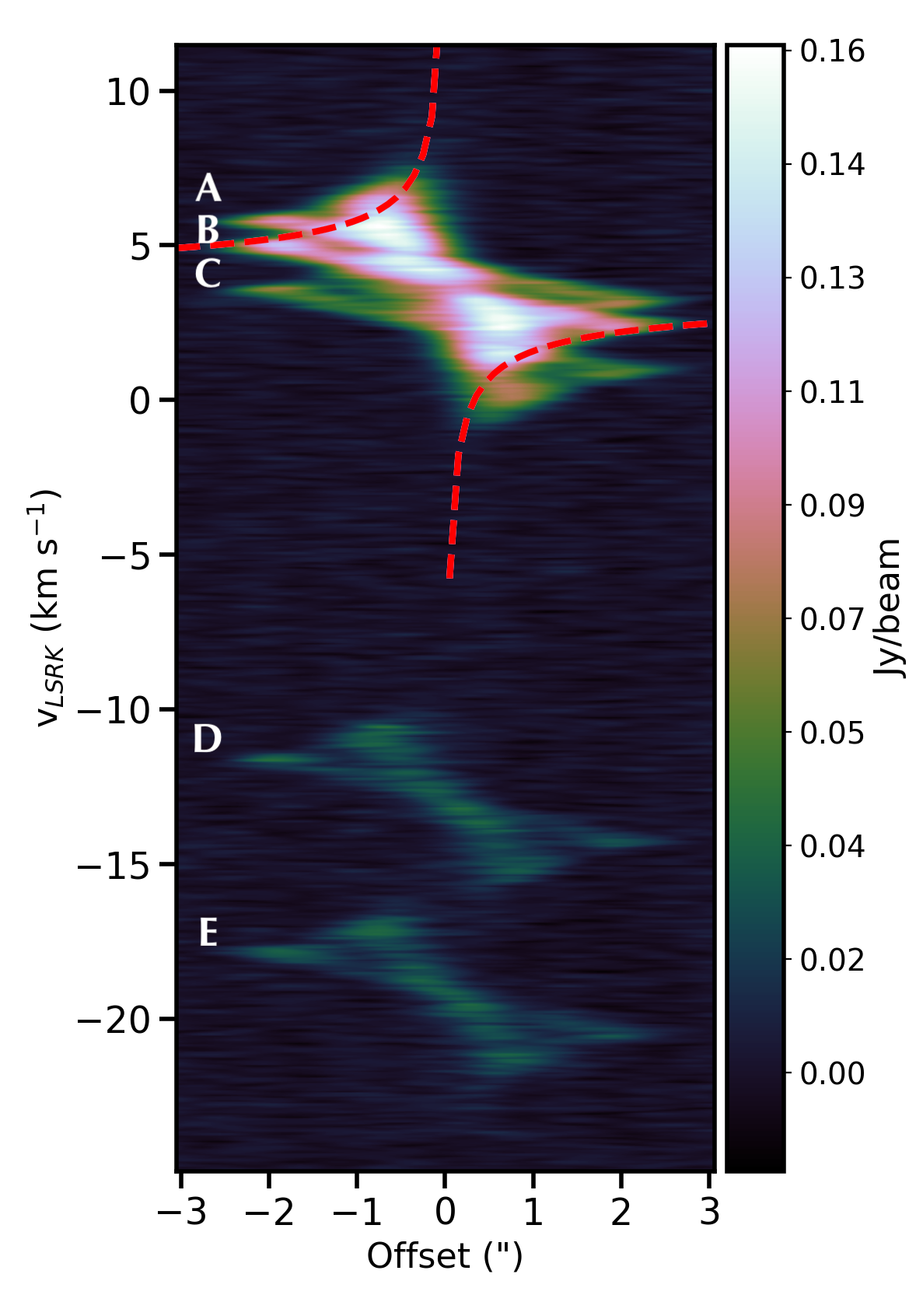}
\caption{Position-velocity diagram along the disk major axis of the CN emission. Dashed curve is a Keplerian  velocity  profile for a stellar mass of 0.61 M$_{\odot}$ at an inclination of 87$^{\rm o}$ and a systemic velocity of 3.70 km s$^{-1}$ LSRK. Labels A, B, C, D and E indicate the 5 observed HF transitions, see Table \ref{Table:Ratios}. 
\label{Fig:PV_DE}}
\end{figure}

The CN N=(2-1) HF components channel maps of the C and D components are displayed in Fig. \ref{Fig:Channel_DE}.  These satellite components also present a typical Keplerian rotation in a disk. In  Figure \ref{Fig:PV_DE}, we present PV diagrams along the disk major axis overlaid with curves representing Keplerian rotation in a geometrically thin disk with an inclination of 87 $\pm$ 1.0 deg, a central mass of 0.60 $\pm$ 0.02 M$_{\odot}$ and v$_{LSRK}$ of 3.70 $\pm$ 0.5 km s$^{-1}$, assuming a distance of 120 pc.

\section{CS and CO emission: Channel maps}
\label{App:B}

\begin{figure*}
\centering
\includegraphics[width=0.9\textwidth]{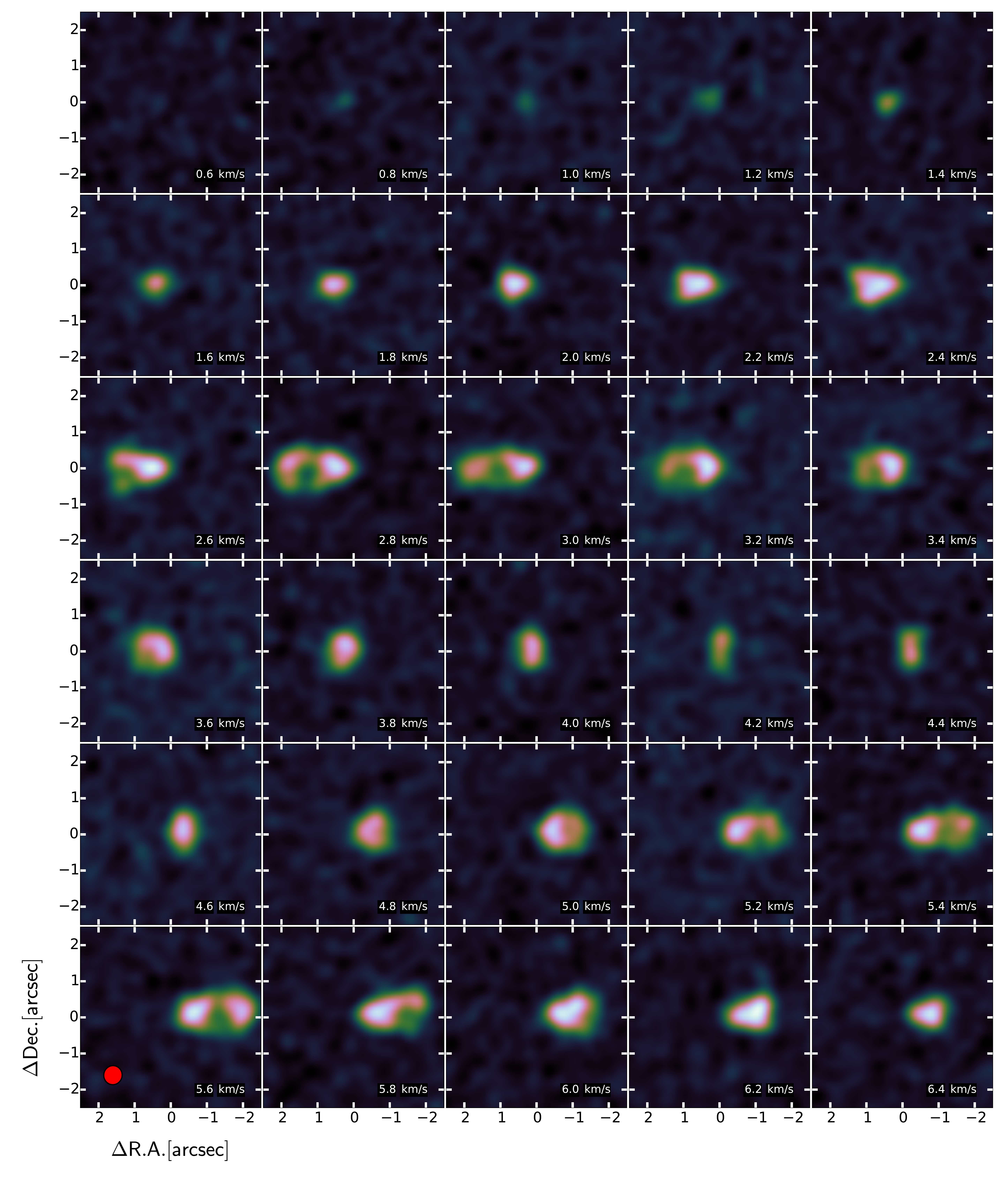}
\caption{Channel maps of the CS emission in the Flying Saucer from 0.60 to 6.40 km s$^{-1}$. The spectral resolution is 0.2 km s$^{-1}$ and the rms per channel is 3.0 mJy beam$^{-1}$.The synthesized beam size is represented in the lower left panel.
\label{Fig:CS_Channel}}
\end{figure*}

\begin{figure*}
\centering
\includegraphics[width=0.9\textwidth]{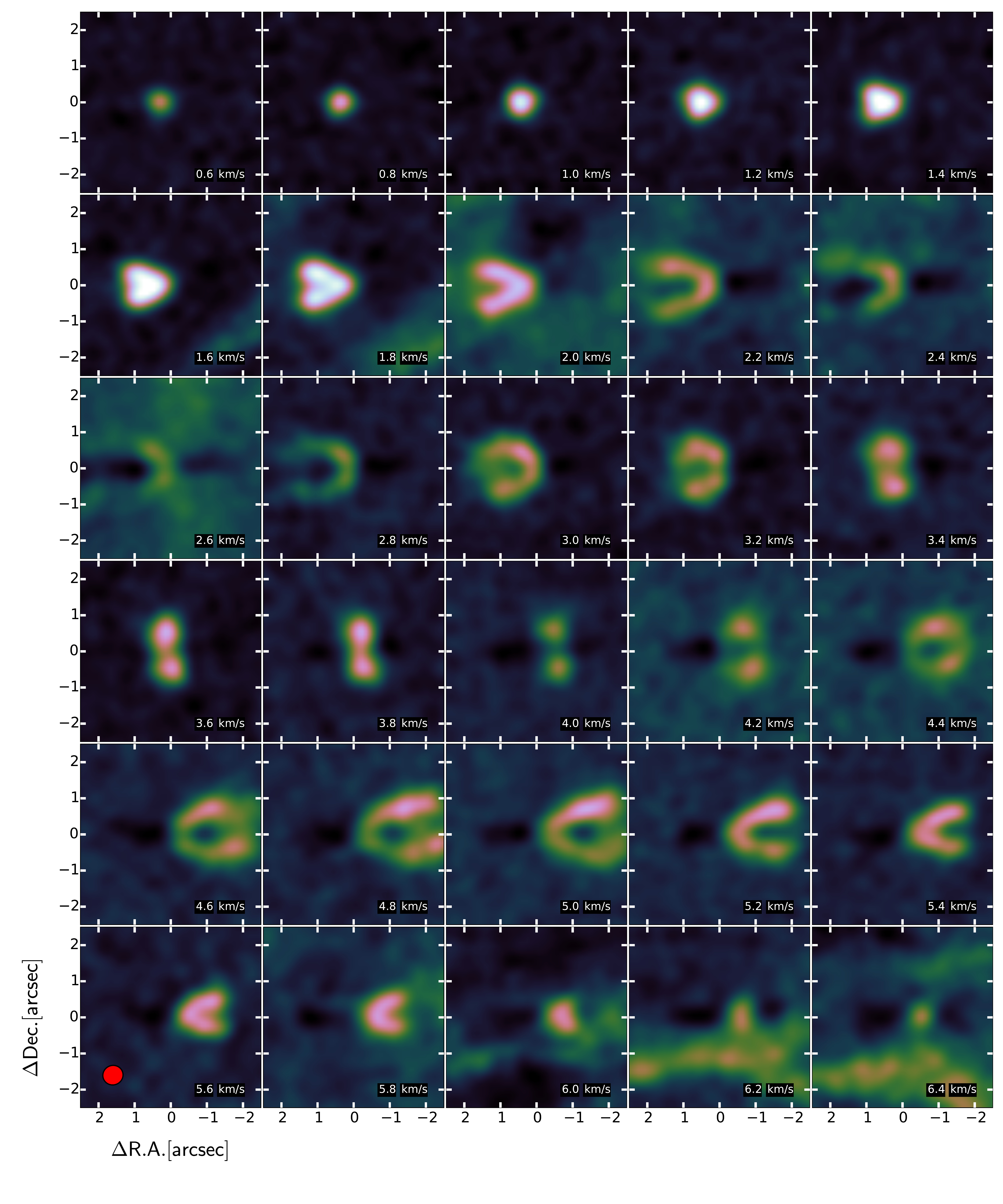}
\caption{Channel maps of the CO emission in the Flying Saucer from 0.60 to 6.40 km s$^{-1}$. The spectral resolution is 0.2 km s$^{-1}$ and the rms per channel is 5.0 mJy beam$^{-1}$.The synthesized beam size is represented in the lower left panel.
\label{Fig:CO_Channel}}
\end{figure*}

Channel maps of CS and CO toward Flying Saucer disk are displayed in Figs. \ref{Fig:CS_Channel} and \ref{Fig:CO_Channel}. The velocity range is similar to channel maps for CN presented in Fig. \ref{Fig:Channel} for a direct comparison.

\end{appendix}

\end{document}